\documentclass[11pt]{article}
\usepackage{comment}
\usepackage[margin=2.5cm]{geometry}
\usepackage{amsmath,amssymb,extarrows,mathtools,graphicx,subfigure,setspace,bbold}
\usepackage{cite}
\usepackage{slashed}
\usepackage[dvipsnames]{xcolor}
\usepackage{hyperref}
\usepackage{physics}
\hypersetup{colorlinks=true, linkcolor=blue, citecolor=magenta, linktoc=page}
\usepackage{tikz}
\usepackage{arydshln,leftidx,mathtools}
\usetikzlibrary{decorations.pathreplacing}
\numberwithin{equation}{section}
\newcommand{\be}{\begin{equation}}
\newcommand{\bea}{\begin{eqnarray}}
\newcommand{\eea}{\end{eqnarray}}
\newcommand{\ba}{\begin{align}}
\newcommand{\ea}{\end{align}}
\newcommand{\ee}{\end{equation}}

\newcommand{\dt}{{\tilde{d}}}
\newcommand{\rft}{{}}
\newcommand{\ct}{{\tilde{c}}}
\newcommand{\Ct}{{\tilde{\mathcal{C}}}}

\usepackage{chngcntr}
\counterwithout*{figure}{section}

\begin{document}

\begin{titlepage}
\vspace{10mm}
\begin{flushright}
IPM/P-2019/007 \\

\end{flushright}

\vspace*{20mm}
\begin{center}

{\Large {\bf Some Aspects of Entanglement Wedge Cross-Section}}

\vspace*{15mm}
\vspace*{1mm}
{Komeil Babaei Velni${}^{\ast}$, M. Reza Mohammadi Mozaffar${}^{\ast, \dagger}$ and M. H. Vahidinia${}^{\ddagger, \dagger}$}

 \vspace*{1cm}

{\it  ${}^\ast$ Department of Physics, University of Guilan,
P.O. Box 41335-1914, Rasht, Iran\\
${}^\dagger$ School of Physics,
Institute for Research in Fundamental Sciences (IPM),\\
P.O. Box 19395-5531, Tehran, Iran\\
${}^\ddagger$ Department of Physics, Institute for Advanced Studies in Basic Sciences (IASBS),\\ P.O. Box  45137-66731,  Zanjan, Iran
}

 \vspace*{0.5cm}
{E-mails: {\tt babaeivelni@guilan.ac.ir, mmohammadi@guilan.ac.ir, vahidinia@iasbs.ac.ir}}%

\vspace*{1cm}
%%\maketitle
\end{center}

\begin{abstract}
We consider the minimal area of the entanglement wedge cross section (EWCS) in Einstein gravity. In the context of holography, it is proposed that this quantity is dual to different information measures, e.g., entanglement of purification, logarithmic negativity and reflected entropy. 
Motivated by these proposals, we examine in detail the low and high temperature corrections to this quantity and show that it obeys the area law even in the finite temperature. We also study EWCS in nonrelativistic field theories with nontrivial Lifshitz and hyperscaling violating exponents. The resultant EWCS is an increasing function of the dynamical exponent due to the enhancement of spatial correlations between subregions for larger values of $z$. We find that EWCS is monotonically decreasing as the hyperscaling violating exponent increases. We also obtain this quantity for an entangling region with singular boundary in a three dimensional field theory and find a universal contribution where the coefficient depends on the central charge. Finally, we verify that for higher dimensional singular regions the corresponding EWCS obeys the area law.
\end{abstract}

\end{titlepage}

\newpage
\tableofcontents
\noindent
\hrulefill
\onehalfspacing
%%%%%%%%%%%%%%%%%%%%%%%%%%%%%%%%
\section{Introduction}

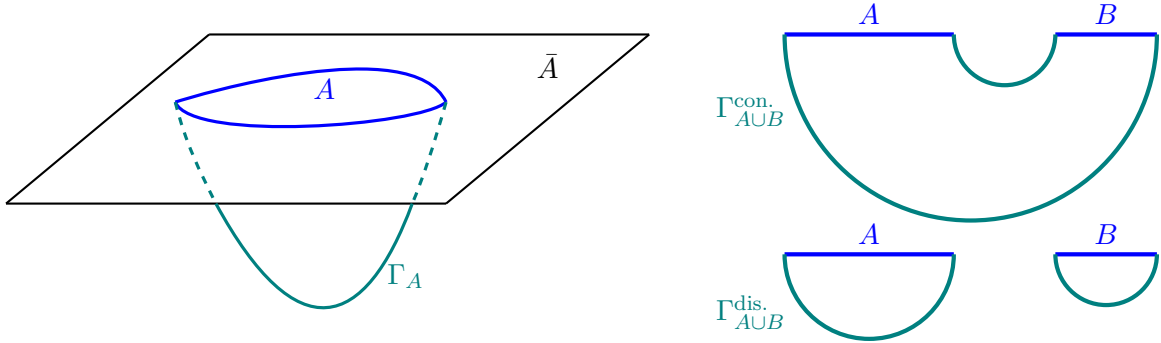
\begin{figure}\label{fig:RT}
\begin{center}
\begin{tikzpicture}[scale=0.9]
\draw[ thick,black] (-3.5,0) -- (3,0);
\draw[ thick,black] (-3.5,0) -- (-0.5,2.5);
\draw[ thick,black] (-0.5,2.5) -- (6,2.5);
\draw[ thick,black] (3,0) -- (6,2.5);

\draw[-,blue,very thick] (-1,1.5) .. controls (-.75,0.9) and (2.6,1.15) .. (3,1.5);

\draw[-,blue,very thick] (-1,1.5) .. controls (-.7,1.6) and (2.5,2.55) .. (3,1.5);

\draw[dashed,teal,very thick] (-1,1.5) .. controls (-0.85,0.9) and (-0.65,0.5) .. (-0.4,0);

\draw[dashed,teal,very thick] (3,1.5) .. controls (2.8,0.8) and (2.65,0.35) .. (2.5,0);
\draw[-,teal,very thick] (-0.4,0) .. controls (0.8,-2.2) and (1.8,-1.9) .. (2.5,0);

\draw[blue] (1.2,1.4) node[above] {$A$}; 

\draw[black] (4.5,1.7) node[above] {$\bar{A}$}; 

\draw[teal] (2.4,-1.4) node[above] {$\Gamma_A$};

\draw[ultra thick,blue] (8,2.5) -- (10.5,2.5);
\draw[ultra thick,blue] (12,2.5) -- (13.5,2.5);
\draw[ultra thick,teal] (10.5,2.5) arc (180:360:0.75cm);
\draw[ultra thick,teal] (8,2.5) arc (180:360:2.75cm);

\draw[ultra thick,blue] (8,-0.75) -- (10.5,-0.75);
\draw[ultra thick,blue] (12,-0.75) -- (13.5,-0.75);
\draw[ultra thick,teal] (8,-0.75) arc (180:360:1.25cm);
\draw[ultra thick,teal] (12,-0.75) arc (180:360:0.75cm);

\draw[blue] (9.25,2.5) node[above] {$A$}; 
\draw[blue] (12.75,2.5) node[above] {$B$}; 

\draw[blue] (9.25,-0.75) node[above] {$A$}; 
\draw[blue] (12.75,-0.75) node[above] {$B$}; 

\draw[teal] (7.5,1) node[above] {$\Gamma_{A\cup B}^{\text{con.}}$}; 
\draw[teal] (7.5,-2) node[above] {$\Gamma_{A\cup B}^{\text{dis.}}$};

%\draw[ thick,blue,dashed] (-2,0) -- (2,0);
%\draw[->, thick,black] (2.1,0) -- (4,0);
%\draw[blue] (-0,0) node[above] {$A$}; 
%\draw[very thick,green] (-2.05,0) arc (180:360:2.08cm);
%\draw[red,very thick] (-2.1,0) .. controls (-0,-4) .. (2.1,0);
%\draw[violet,very thick] (-2.1,0) .. controls (-0,-1) .. (2.1,0);
%
%\draw[] (-2.15,0) node[above] {$-\frac{\ell}{2}$}; 
%\draw[] (2.05,0) node[above] {$\frac{\ell}{2}$}; 
%\draw[] (-5.5,1) node[left] {$\rho=\epsilon$}; 
%\draw[] (-4,-2.2) node[left] {$\rho$}; 
%\draw[] (4,0) node[above] {$x$}; 
%%\draw[] (5.5,1) node[right] {bondary of $AdS_3$ where $CFT_2$ lives}; 
%\draw[->, thick,black] (-3.9,-1.6) -- (-3.9,-2.5);
%%\draw[->,brown,very thick] (-3.5,0) .. controls (-3,2) and (-2.6,0.5) .. (-1.5,1);
%\draw[->,brown,very thick] (-3.5,0) .. controls (-4,2) and (-4.4,0.5) .. (-5.5,1);
\end{tikzpicture}
\caption{Schematic configurations for computing $S_A$ (left) and $S_{A\cup B}$ (right). Note that in the latter case we have two different extremal configurations denoted by $\Gamma_{A\cup B}^{\text{con.}}$ and $\Gamma_{A\cup B}^{\text{dis.}}$ corresponding to connected and disconnected RT surfaces respectively.}
\end{center}
\end{figure}

\begin{figure}
\begin{center}
\includegraphics[scale=1]{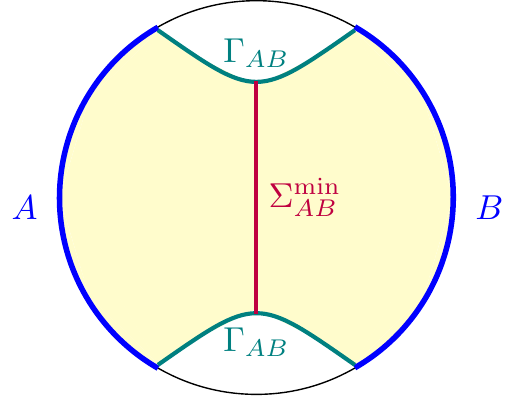}
\hspace*{1cm}
\includegraphics[scale=1]{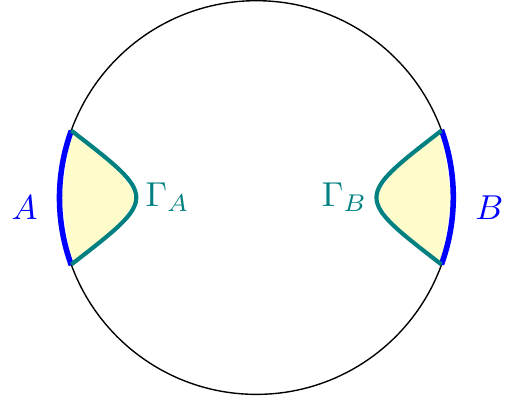}
\end{center}
\caption{\textit{Left}: Schematic configuration for computing $E_W$ where the entanglement wedge (shaded region) is connected. In this case $E_W$ is proportional to the area of $\Sigma_{AB}^{\rm min}$. \textit{Right}: For small entangling regions where the entanglement wedge is disconnected and $\Sigma_{AB}^{\rm min}$ becomes empty the corresponding $E_W$ vanishes.}
\label{fig:eopregions}
\end{figure}

In recent years, the study of quantum information concepts such as entanglement using gauge/gravity correspondence has been an active area of research. In particular, improving the duality by adding an explicit relation between a measure of entanglement in the boundary theory and a geometric entity which lives in the bulk spacetime is of great interest. In this context, the Ryu-Takayanagi (RT) proposal is a remarkably simple prescription to compute entanglement entropy (EE) for the QFTs dual to Einstein gravity. Consider a spatial region $A$ in the boundary field theory, the corresponding entanglement entropy between $A$ and its complement is given by\cite{Ryu:2006bv}
\bea
S_A=\frac{{\rm area}(\Gamma_A)}{4G_N},
\eea
where $G_N$ is the Newton constant and $\Gamma_A$ is a codimension-2, spacelike minimal hypersurface in the bulk spacetime, anchored
to the asymptotic boundary such that $\partial \Gamma_A=\partial A$ (see figure \ref{fig:RT}). 
The RT proposal which passes a variety of consistency tests, generalizes to time dependent case \cite{Hubeny:2007xt} and higher derivative theories of gravity \cite{Hung:2011xb,Dong:2013qoa,Camps:2013zua,Mozaffar:2016hmg}.
Using these prescriptions, the correlation of several disconnected components can also be considered. In particular when the entangling region is made by two disjoint spatial components, an important quantity to study is the holographic mutual information (HMI) given as follows\footnote{In the following we denote $S_{A\cup B}$ by $S_{AB}$.}
\bea\label{HMI}
I(A, B)=S_A+S_B-S_{A\cup B}.
\eea
The mutual information is free from UV divergences and subadditivity guarantees that $I(A,B)\geq 0$. In \cite{Headrick:2010zt} it was shown that HMI exhibits a phase transition which is due to the competition between two different configurations for computing $S_{AB}$. Indeed, in order to find this contribution we should consider two minima corresponding to a connected configuration and to a disconnected one (see figure \ref{fig:RT}). At small distances the connected configuration has the minimal area, while for large separations the RT surface changes topology and the disconnected configuration is favored. Hence using eq.\eqref{HMI} the HMI vanishes in the latter case. It is worth to mention that HMI phase transition is a feature of large $N$ limit of quantum field theory and considering $\mathcal{O}(\frac{1}{N})$ corrections changes this picture\cite{Faulkner:2013ana}.
Besides the already mentioned case of the
HEE and HMI, there are many attempts to construct a holographic prescription for other information measures, e.g., relative entropy\cite{Jafferis:2015del}, quantum information metric\cite{MIyaji:2015mia} and computational complexity\cite{Susskind:2014rva,Brown:2015bva}. However, in this paper, we focus on another concept that has recently entered this discussion which is the minimal area of the entanglement wedge cross section (EWCS) and on its conjectured holographic duals\cite{Takayanagi:2017knl,Nguyen:2017yqw,Kudler-Flam:2018qjo,Tamaoka:2018ned,Dutta:2019gen}. Given a particular
spatial region composed of two components $A$ and $B$ in the boundary theory, the EWCS is given as follows\footnote{In the following $E_W$ is used interchangeably with EWCS.}
\bea\label{heop}
E_W(\rho_{AB})=\frac{{\rm area}(\Sigma_{AB}^{\min})}{4G_N},
%E_P(\rho_{AB})=\min_{\Gamma^{(A)}_{AB}\subset \Gamma^{\rm min}_{AB}}\frac{{\rm area}(\Sigma_{AB}^{\min})}{4G_N},
\eea
where $\Sigma_{AB}^{\min}$ is the minimal cross sectional area of the corresponding entanglement wedge (see Fig.\ref{fig:eopregions}). Let us now briefly mention different holographic interpretation of this object. 

EWCS was proposed by \cite{Takayanagi:2017knl,Nguyen:2017yqw} to be dual to the entanglement
of purification (EoP), but there is no proof and the conjecture is mainly based on some properties that EoP should satisfy. 
Using the entanglement negativity in quantum error-correcting codes and tensor network models of holography, another proposal has been made for $E_W$ in \cite{Kudler-Flam:2018qjo}. The authors of this paper identified properties of EWCS with entanglement
negativity in holographic theories and made explicit comparisons between $E_W$ and entanglement negativity in a 2-dimensional CFT. Also recently another interpretation of EWCS is given in \cite{Dutta:2019gen} where the authors show that the entanglement entropy associated to a canonical purification is captured by $E_W$. They call this quantity the reflected entropy, i.e., $S_R$, which is a
measure of quantum and classical correlations between $A$ and $B$.

In the following we will focus on the first proposal given in \cite{Takayanagi:2017knl,Nguyen:2017yqw} which gives entanglement of purification  in terms of $E_W$. Consider a bipartite system with Hilbert space equal to the direct product of two
factors, i.e., $\mathcal{H}=\mathcal{H}_A\otimes \mathcal{H}_B$ and let $\rho_{AB}$ be a density matrix corresponding to a mixed state on $\mathcal{H}$. It is a well known fact that by adding auxiliary degrees of freedom to $\mathcal{H}$ one can construct a pure state $|\psi\rangle$ out of $\rho_{AB}$ such that $\rho_{AB}={\rm tr}_{A'B'}|\psi\rangle\langle\psi|$ and $|\psi\rangle\in \mathcal{H}_{AA'}\otimes \mathcal{H}_{BB'}$. Indeed, this procedure is not unique and one can find different purifications for a given mixed state. Now following \cite{Terhal:2002}, the EoP is defined as
\bea
E_P(\rho_{AB})=\min_{\rho_{AB}={\rm tr}_{A'B'}|\psi\rangle\langle\psi| }S_{\rho_{AA'}},
\eea
where $\rho_{AA'}={\rm tr}_{BB'}|\psi\rangle\langle\psi|$ and the minimization is taken over any $|\psi\rangle$. The EoP is a measure of correlation between $A$ and $B$ and reduces to EE for pure states. Considering a general quantum system, the EoP
is subject to the following inequalities\footnote{For a complete set of inequalities see \cite{{Terhal:2002},{Bagchi:1502}} and appendix A of \cite{Nguyen:2017yqw}.}
\bea\label{inequality}
\frac{I(A, B)}{2}\leq E_P(\rho_{AB})\leq {\rm min}\left(S_A, S_B\right),\nonumber\\
E_P(\rho_{A(BC)})\geq \frac{I(A, B)}{2}+\frac{I(A, C)}{2}.
\eea
Based on \cite{Takayanagi:2017knl,Nguyen:2017yqw} the holographic EoP of the boundary field theory is given by $E_P(\rho_{AB})=E_W(\rho_{AB})$. Using this prescription, it was shown that, the resultant quantity obeys all the properties of EoP. Also it was shown that, keeping the geometry of $A$ and $B$ fixed while their separation increases, the holographic EoP has a phase transition such that $E_P=0$ when the two regions are distant enough. This behavior which is similar to the phase transition of HMI is due to the competition between
two different configurations for the entanglement wedge. Despite the fact that in large distance limit, HMI vanishes continuously, the EoP experiences a discontinuous transition. In \cite{Bhattacharyya:2018sbw}, EoP for a 2-dimensional scalar theory was studied where its behavior qualitatively agrees with the conjectured holographic proposal. Also using a generalization of the above proposal to time dependent backgrounds, \cite{Nguyen:2017yqw,Yang:2018gfq} studied the evolution of EoP after a quantum quench in the dual field theory. Related investigations attempting to better understand EoP both in the field theory and holography have also appeared in \cite{Bao:2017nhh,Hirai:2018jwy,Espindola:2018ozt,Bao:2018gck,Umemoto:2018jpc,Bao:2018fso,Agon:2018lwq,Caputa:2018xuf,Liu:2019qje,Bhattacharyya:2019tsi,Ghodrati:2019hnn}.

The aim of this paper is to more investigate the holographic aspects of EWCS in field theories dual to Einstein gravity. We will study the phase transition of EWCS in a relativistic theory at finite temperature and find the low and high temperature expansion of this quantity. We show explicitly that EWCS obeys an area law scaling even in finite temperature. We also study the transition of EWCS in a non-relativistic QFT with nontrivial dynamical and hyperscaling violating exponents, i.e., $z$ and $\theta$. Moreover, we investigate the properties of EWCS for nonsmooth entangling regions where the boundary contains conical singularity. Considering a simple configuration for the subregions, we find a universal contribution to EWCS due to the presence of corner in a four dimensional field theory when the subregions coincide.

This paper is organized as follows. In section \ref{sec:finiteT}, after a short review on HEE and HMI for a strip entangling region at finite temperature, we investigate the corresponding
 EWCS in different dimensions and obtain analytical results at low and high temperature limits. The role of dynamical and hyperscaling violating exponents in the phase transition of EWCS is discussed in section \ref{sec:nonrel}. In section \ref{sec:eopcorner}, we study the corner contributions to EWCS considering a union of kinks and creases in four and higher dimensions.  We conclude with a discussion of our results, as well as possible future directions, in section \ref{sec:conclusions}. We relegate some details of the computations to the appendix.

%\begin{figure}\label{0}
%\begin{center}
%\begin{tikzpicture}[scale=.5]
%\draw[ultra thick,blue] (-1,0) -- (2,0);
%\draw[ultra thick,blue] (4,0) -- (6,0);
%\draw[ultra thick,teal] (2,0) arc (180:360:1cm);
%\draw[ultra thick,teal] (-1,0) arc (180:360:3.5cm);
%
%
%
%
%\draw[ultra thick,blue] (13,0) -- (16,0);
%\draw[ultra thick,blue] (18,0) -- (20,0);
%\draw[ultra thick,teal] (13,0) arc (180:360:1.5cm);
%\draw[ultra thick,teal] (18,0) arc (180:360:1cm);
%
%\draw[] (0.5,0) node[above] {$A$}; 
%\draw[] (5,0) node[above] {$B$}; 
%
%\draw[] (14.5,0) node[above] {$A$}; 
%\draw[] (19,0) node[above] {$B$}; 
%
%\draw[teal] (-3,-1) node[above] {$\Gamma_{A\cup B}^{\text{con.}}$}; 
%\draw[teal] (11,-1) node[above] {$\Gamma_{A\cup B}^{\text{dis.}}$}; 
%
%\end{tikzpicture}
%\caption{Schematic figure of the two different configurations for computing $S(A_1\cup A_2)$. }
%\end{center}
%\end{figure}
%

\section{EWCS at Finite Temperature in Relativistic Theories}\label{sec:finiteT}

In this section we study the finite temperature contribution to the $E_W$ for holographic
theories dual to Einstein gravity. We begin by reviewing the calculation
of the finite temperature corrections to HEE and HMI using a systematic expansion, which was originally performed in \cite{Fischler:2012ca, Fischler:2012uv}.\footnote{On the CFT side thermal corrections to EE is computed in \cite{Cardy:2014jwa,Herzog:2014fra}.} Then applying
this method allows us to evaluate the thermal corrections to $E_W$ for a straight belt entangling region in section \ref{sec:EoPTLH}.

\begin{figure}
\begin{center}
\includegraphics[scale=1]{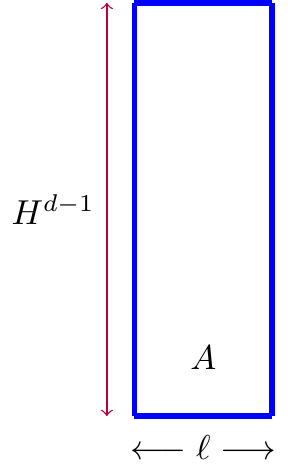}
\hspace*{1cm}
\includegraphics[scale=1]{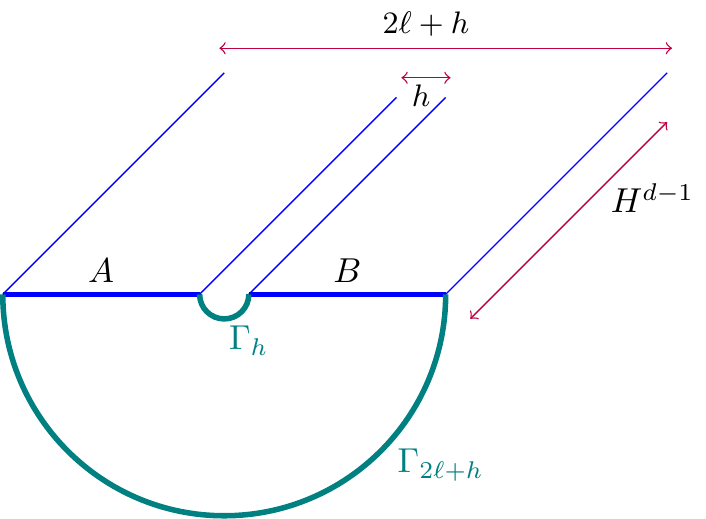}
\end{center}
\caption{Schematic configuration for computing HEE (left) and HMI (right). In the latter case, we just demonstrate the connected configuration where the HMI is non-zero.}
\label{fig:regions}
\end{figure}

The bulk geometry will be a $(d+2)$-dimensional AdS black brane in Poincare coordinates
\bea\label{metric}
ds^2=\frac{L^2}{r^2}\left(-f(r)dt^2+\frac{dr^2}{f(r)}+d\vec{x}^2\right),\;\;\;f(r)=1-\frac{r^{d+1}}{r_0^{d+1}},
\eea
where $r_0$ is the horizon radius and $L$ is the AdS radius. In the following without loss of generality we set  $L=1$. 
From eq.\eqref{metric}, one obtains that the temperature and thermal entropy density are given by
\bea\label{temp}
T=\frac{d+1}{4\pi r_0},\;\;\;\;\;\;\; s_{\rm th}=\frac{1}{4G_N}\frac{1}{r_0^d}.
\eea
Figure \ref{fig:regions} shows the entangling regions that we consider for computing HEE and HMI. The straight belt entangling region can be parametrized as 
\bea\label{stripregion}
-\frac{\ell}{2}\leq x_1(r)\equiv x(r)\leq\frac{\ell}{2},\;\;\;\;-\frac{H}{2}\leq x_i\leq\frac{H}{2},\;\;\;i=2,\cdots,d,
\eea
where we assume $H\gg\ell$.

%\subsection{EoP at zero Temperature}\label{EoPT0}
%
%The profile of the minimal hypersurface has two branches
%\bea
%x_{\pm}(r)=\pm\frac{\sqrt{\pi}}{2}\frac{\Gamma(\frac{d+1}{2d})}{\Gamma(\frac{1}{2d})}r_t\mp \frac{r}{d+1}\left(\frac{r}{r_t}\right)^{d}{_2}F_1\left(\frac{1}{2},\frac{d+1}{2d},\frac{3d+1}{2d},\left(\frac{r}{r_t}\right)^{2d}\right),
%\eea
%where the length of the entangling region is given by
%\bea
%\frac{\ell}{2}=x_+(0)-x_-(0)=\sqrt{\pi}\frac{\Gamma(\frac{d+1}{2d})}{\Gamma(\frac{1}{2d})}r_t.
%\eea

\subsection{Low and High Temperature Behavior of HEE and HMI}\label{sec:LHT}

In this section we review the computation of HEE for strip entangling region in the low and high temperature limit. This analysis has been done in \cite{Fischler:2012ca} for HEE and generalized to HMI in \cite{Fischler:2012uv}. Here we present the main steps and fix our notations.\footnote{The following analysis does not apply to three dimensional bulk geometry, i.e, $d=1$. We will come back
to this case in section \ref{sec:threedim}.}

Employing the RT prescription and using eq.\eqref{metric}, the corresponding HEE functional is given by
\bea\label{RTstrip}
S=\frac{H^{d-1}}{4G_N}\int\frac{dr}{r^d}\sqrt{\frac{1}{f(r)}+x'(r)^2}.
\eea
Extremizing the above expression yields the equation of motion for $x(r)$, however, since there is no explicit $x(r)$ dependence, the corresponding momentum is a conserved quantity. Therefore we find the following first integral
\bea\label{eomtemp}
x'(r)=\pm\frac{1}{\sqrt{f(r)\left(\left(\frac{r_t}{r}\right)^{2d}-1\right)}},
\eea
where $r_t$ is the turning point of the minimal hypersurface. Using the above expression the relation between $\ell$ and $r_t$ is given by
\bea\label{lengthtemp}
\ell=2r_t\int_0^1\frac{u^d du}{\sqrt{1-u^{2d}}}\left(1-\left(\frac{r_t}{r_0}\right)^{d+1} u^{d+1}\right)^{-\frac{1}{2}}.
\eea
On the other hand plugging eq.\eqref{eomtemp} back into eq.\eqref{RTstrip}, we find
\bea\label{heetemp}
S=\frac{1}{2G_N}\frac{H^{d-1}}{r_t^{d-1}}\int_{\frac{\epsilon}{r_t}}^{1}\frac{du}{u^d\sqrt{1-u^{2d}}}\left(1-\left(\frac{r_t}{r_0}\right)^{d+1} u^{d+1}\right)^{-\frac{1}{2}}.
\eea

The above integrals can be carried out analytically in $d=1$ case. Hence before going to general dimensions, let us first consider
this special case which corresponds to a three dimensional bulk geometry.

%For definiteness, we consider the entanglement
%of purification, and show that its dependence exhibits the
%different regimes discussed in the introduction.

\subsubsection{HEE and HMI in $d=1$}\label{d=1}

For $d=1$ the minimal surface is a spacelike geodesic whose length can be expressed analytically in closed form \cite{Ryu:2006ef}, which enables us to directly extract its temperature behavior in various regimes. In this case eqs.\eqref{lengthtemp} and \eqref{heetemp} become
\bea
\ell=2r_t\int_0^1\frac{u du}{\sqrt{1-u^{2}}}\left(1-\left(\frac{r_t}{r_0}\right)^{2} u^{2}\right)^{-\frac{1}{2}},\;\;\;
S=\frac{1}{2G_N}\int_{\frac{\epsilon}{r_t}}^{1}\frac{du}{u\sqrt{1-u^{2}}}\left(1-\left(\frac{r_t}{r_0}\right)^{2} u^{2}\right)^{-\frac{1}{2}}.
\eea
It is straightforward to evaluate these quantities and to produce the result
\bea\label{lengthheed1}
\ell=r_0\log\frac{r_0+r_t}{r_0-r_t},\;\;\;S=\frac{c}{3}\log\left(\frac{1}{\pi\epsilon T}\sinh \pi\ell T\right),
\eea
where $c=\frac{3}{2G_N}$ denotes the central charge of dual two dimensional CFT. We can make use of the above expression to find the low and high temperature behavior of HEE as follows
\bea
S\sim S_{\rm div.}+\Bigg\{ \begin{array}{rcl}
&\frac{c}{18}(\pi \ell T)^2+\cdots &\,\,\,\ell T\ll 1\\
&\frac{c}{3}\pi \ell T+\cdots &\,\,\,\ell T\gg 1
\end{array}.
\eea
This shows that the thermal fluctuations increases HEE, as expected. In particular, in high temperature limit the leading finite term in HEE takes precisely the form expected for the volume law contribution to the entanglement entropy in the dual field theory due to thermal fluctuations. That is, the leading thermal contribution is proportional to $\ell$. 

In order to investigate the low and high temperature behavior of HMI, one should keep in mind that we have three different scales, i.e., $h$, $\ell$ and $T$. Considering low temperature with respect to the subregion sizes and the separation between them corresponds to $h T\ll \ell T\ll 1$. One might also regard the $h T\ll 1 \ll \ell T$ case where we only have low temperature with respect to the separation scale. As demonstrated in \cite{Fischler:2012uv} these two different limits contain distinct physics. Further, in the following we are only interested in cases where HMI is non-zero so we neglect the $hT\gg\ell T$ or $1\ll hT\ll\ell T$ cases which correspond to disconnected configurations for RT surfaces with zero HMI. For the connected configuration eq.\eqref{HMI} becomes
\bea\label{HMI1}
I=2S(\ell)-S(h)-S(2\ell+h).
\eea
Using eq.\eqref{lengthheed1} evaluating the above expression is a straightforward exercise,
which yields
\bea\label{hmitempd1}
I=\frac{c}{3}\log\frac{\sinh^2(\pi\ell T)}{\sinh(\pi h T)\;\sinh(\pi(2\ell+h) T)}.
\eea
Equipped with the above result we can compute HMI in different scaling regimes as follows
\bea
I\sim \frac{c}{3}\Bigg\{ \begin{array}{rcl}
&\log\frac{\ell}{2h}-\frac{1}{3}(\pi\ell T)^2+\cdots&\,\,\,hT\ll\ell T\ll 1\\
&-\log(2\pi hT)-\pi hT+\log\tanh(\pi \ell T)+\cdots &\,\,\,hT\ll 1 \ll \ell T
\end{array},
\eea
which demonstrates that HMI is a monotonically decreasing function of temperature. Taking the limit for adjacent subregions  $h\rightarrow 0$ in the above result, we see that the HMI diverges.

\subsubsection{HEE and HMI in $d>1$}
While the integral in eqs.\eqref{lengthtemp} and \eqref{heetemp} cannot be carried out analytically for general $d>1$, to perform an exact estimation, we employ a particular series expansion which is enough to extract the main behavior of HEE at finite temperature. Using this expansion eq.\eqref{lengthtemp} can be written as follows (see \cite{Fischler:2012ca} for details)
\bea\label{ellrt}
\ell=2r_t\sum_{n=0}^{\infty}\frac{1}{1+n(d+1)}\frac{\Gamma\left(n+\frac{1}{2}\right)}{\Gamma\left(n+1\right)}\frac{\Gamma\left(\frac{(d+1)(n+1)}{2d}\right)}{\Gamma\left(\frac{(d+1)n+1}{2d}\right)}\left(\frac{r_t}{r_0}\right)^{n(d+1)},
\eea
which converges for $r_t<r_0$. On the other hand using similar expansion in eq.\eqref{heetemp}, we find
\bea\label{heeddim}
S=\frac{1}{2(d-1)G_N}\left(\frac{H}{\epsilon}\right)^{d-1}+S_{\rm finite.},
\eea
where
\bea\label{Sfinite}
S_{\rm finite.}=\frac{1}{4G_N}\frac{H^{d-1}}{r_t^{d-1}}\left(\frac{c}{1-d}+\sum_{n=1}^{\infty}\frac{1}{d}\frac{\Gamma\left(n+\frac{1}{2}\right)}{\Gamma\left(n+1\right)}\frac{\Gamma\left(\frac{n(d+1)-d+1}{2d}\right)}{\Gamma\left(\frac{n(d+1)+1}{2d}\right)}\left(\frac{r_t}{r_0}\right)^{n(d+1)}\right),
\eea
and we have define $c=\frac{2\sqrt{\pi}\Gamma \left(\frac{d+1}{2d}\right)}{\Gamma \left(\frac{1}{2d}\right)}$. 
Now it is straightforward to find the low and high temperature behavior of HEE using eqs. \eqref{ellrt} and \eqref{Sfinite}. 

\subsubsection*{(i) HEE at Low Temperature Limit $\ell T\ll 1$}
In this case using eq.\eqref{temp} the $\ell T\ll 1$ limit can be interpreted in terms of bulk data as $r_t\ll r_0$. In this limit eq.\eqref{ellrt} yields 
\bea\label{rtsmalltemp}
r_t=\frac{\ell}{c}\left(1-\frac{\sqrt{\pi}}{(d+2)c^{d+2}}\frac{\Gamma\left(\frac{d+1}{d}\right)}{\Gamma\left(\frac{d+2}{2d}\right)}\left(\frac{\ell}{r_0}\right)^{d+1}+\mathcal{O}\left(\left(\frac{\ell}{r_0}\right)^{2(d+1)}\right)\right).
\eea
Substituting the above expression into eq.\eqref{Sfinite} and expand to leading order in the temperature, we finally obtain
\bea\label{sfinitelowtemp}
S_{\rm finite.}=\frac{\mathcal{C}_0}{4G_N} \frac{H^{d-1}}{\ell^{d-1}}\left(1+\mathcal{C}_1(\ell T)^{d+1}+\mathcal{O}((\ell T)^{2(d+1)})\right),
\eea
where 
\bea
\mathcal{C}_0= \frac{c^d}{1-d},\;\;\;\;\;\;
\mathcal{C}_1=\left(\frac{4\pi}{d+1}\right)^{d+1}\frac{\sqrt{\pi}}{2c^{d+2}}\frac{1-d}{d+2}\frac{\Gamma \left(\frac{1}{d}\right)}{\Gamma \left(\frac{d+2}{2d}\right)}.
\eea
According to eq.\eqref{sfinitelowtemp}, the thermal fluctuations increases the HEE, as expected (note that $\mathcal{C}_0$ and $\mathcal{C}_1$ are both negative). 

\subsubsection*{(ii) HEE at High Temperature Limit $\ell T\gg 1$}
As the length of the subregion becomes large, the turning point of the corresponding hypersurface approaches the horizon and eventually, the minimal hypersurface covers a part of the horizon. In this case the entanglement entropy is determined entirely by the contributions coming from the near horizon part of the minimal hypersurface\cite{Ryu:2006ef}. Hence we must consider eqs.\eqref{ellrt} and \eqref{Sfinite} in the limit that $r_t \rightarrow r_0$. After some algebra, eq.\eqref{Sfinite} becomes
\bea
S_{\rm finite}=\frac{H^{d-1}}{4G_Nr_t^{d-1}}\left(\frac{\ell}{r_t}-\frac{dc}{d-1}
+\sum_{n=1}^{\infty}\frac{2d}{n(d+1)+1-d}\frac{1}{n(d+1)+1}\frac{\Gamma \left(n+\frac{1}{2}\right)}{\Gamma (n+1)}\frac{\Gamma\left(\frac{(d+1)(n+1)}{2d}\right)}{\Gamma\left(\frac{n(d+1)+1}{2d}\right)}\left(\frac{r_t}{r_0}\right)^{n(d+1)}\right).\nonumber
\eea
It is easy to see that in the large $n$ limit the above infinite series behaves as $\frac{1}{n^2}\left(\frac{r_t}{r_0}\right)^{n(d+1)}$, so we can safely consider $r_t\rightarrow r_0$ limit. Hence the final result for the finite part of the HEE in this case becomes
\bea\label{sfinitehightemp}
S_{\rm finite}=\frac{\ell H^{d-1}}{4G_N}\left(\frac{4\pi T}{d+1}\right)^d\left(1+\frac{d+1}{4\pi \ell T}\mathcal{C}_2\right),
\eea
where
\bea
\mathcal{C}_2=-\frac{dc}{d-1}+2\sum_{n=1}^{\infty}\frac{d}{n(d+1)+1-d}\frac{1}{n(d+1)+1}\frac{\Gamma \left(n+\frac{1}{2}\right)}{\Gamma (n+1)}\frac{\Gamma\left(\frac{(d+1)(n+1)}{2d}\right)}{\Gamma\left(\frac{n(d+1)+1}{2d}\right)}.
\eea
Note that the first term in eq.\eqref{sfinitehightemp} shows a volume law which is a typical property of entanglement entropy at finite temperature and shows that for a mixed thermal state EE measures both classical and quantum correlations.

\subsubsection*{(iii) HMI at Low and High Temperature Limit}
Now we are equipped with all we need to calculate the HMI for configuration depicted in figure \ref{fig:regions} using eqs. \eqref{sfinitelowtemp} and \eqref{sfinitehightemp}. As we mentioned before, we are only interested in cases where the HMI is non-zero corresponding to a connected configuration. In this case using eqs. \eqref{HMI} and \eqref{heeddim}, we have
\bea\label{HMId}
I=2S_{\rm finite.}(\ell)-S_{\rm finite.}(h)-S_{\rm finite.}(2\ell+h).
\eea
Considering the $h T\ll \ell T\ll 1$ limit and using eq.\eqref{sfinitelowtemp} for all three distinct entropies appear in the above expression, yields\cite{Fischler:2012uv,Alishahiha:2014jxa}

\bea\label{hmiexpand1}
I=I_{T=0}-\mathcal{C}_0\mathcal{C}_1\frac{H^{d-1}}{2G_N} (\ell+h)^2T^{d+1},
\eea
where $I_{T=0}$ is the HMI at zero temperature, which in this case is given by 
\bea
I_{T=0}=\mathcal{C}_0\frac{ H^{d-1}}{4G_N}\left(\frac{2}{\ell^{d-1}}-\frac{1}{h^{d-1}}-\frac{1}{(2\ell+h)^{d-1}}\right).
\eea
We note again that the finite temperature corrections reduce the HMI so the mutual correlations between subregions decrease. Another interesting case to consider is $h T\ll 1\ll \ell T$, for which using eqs. \eqref{sfinitelowtemp} and \eqref{sfinitehightemp} we find  
\bea\label{hmiexpand2}
I=\frac{ H^{d-1}T^{d-1}}{4G_N}\left(-\frac{\mathcal{C}_0}{(hT)^{d-1}}+\left(\frac{4\pi}{d+1}\right)^{d-1}\mathcal{C}_2-\left(\frac{4\pi}{d+1}\right)^dhT-\mathcal{C}_0\mathcal{C}_1(hT)^2\right).
\eea
The second term in the above expression which is proportional to the area of the entangling region shows that the HMI obeys an area law even in finite temperature. 

%Once again we see that the HMI decreases with increasing temperature, as expected.

\subsection{Low and High Temperature Behavior of EWCS}\label{sec:EoPTLH}
\begin{figure}
\begin{center}
\includegraphics[scale=1.3]{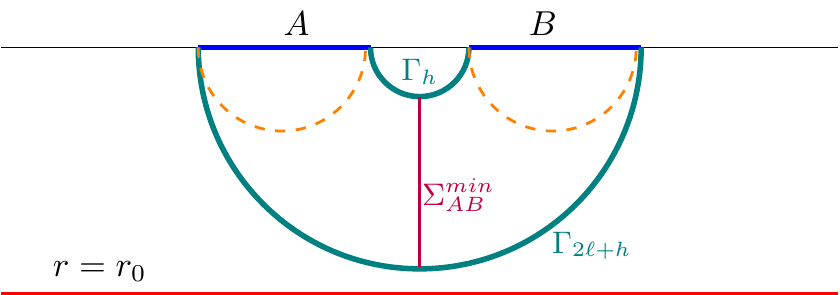}
\end{center}
\caption{Schematic configuration for computing $E_W(\rho_{AB})$.}
\label{fig:regionseop}
\end{figure}

We turn now to the calculation of $E_W$, using the recent holographic prescription given in \cite{Takayanagi:2017knl,Nguyen:2017yqw}. According to this, the $E_W$ of a certain combined region $AB$ is given by eq.\eqref{heop}. Since we focus on the case of two intervals which have a reflection symmetry about $x=0$, we expect that the corresponding minimal configuration respects this symmetry. Indeed, in this case $\Sigma_{AB}^{\rm min}$ runs along the radial direction and connects the corresponding turning points of $\Gamma_h$ and $\Gamma_{2\ell+h}$(see Fig.\ref{fig:regionseop}). 
Using eq.\eqref{metric}, the area of this hypersurface can be written as 
\bea\label{eopfunc}
E_W=\frac{H^{d-1}}{4G_N}\int_{r_t{(h)}}^{r_t{(2\ell+h)}}\frac{dr}{r^{d}\sqrt{1-\frac{r^{d+1}}{r_0^{d+1}}}}.
\eea
Evaluating the above integral is a straightforward exercise both for $d=1$ and $d>1$. Below we consider these cases separately.

%\bea
%E_P\sim \frac{1}{4G_N}\Bigg\{ \begin{array}{rcl}
%&\frac{c}{6}\log\frac{\tanh\frac{\pi(2\ell+h)T}{2}}{\tanh\frac{\pi hT}{2}},&d=1,\\
%& & \\
%&\frac{H^{d-1}}{1-d}\left(\frac{\sqrt{f(r)}}{r^{d-1}}-\frac{d-3}{4}\frac{r^2}{r0^{d+1}} _2F_1\left(\frac{1}{2},\frac{2}{d+1},\frac{d+3}{d+1},\frac{r^{d+1}}{r_0^{d+1}}\right)\right), &d>1,
%\end{array},
%\eea

\subsubsection{EWCS in $d=1$}\label{sec:threedim}
In this case where the boundary theory lives in two dimensions, performing the integral in eq.\eqref{eopfunc}, we are left with
\bea
E_W=\frac{1}{4G_N}\log\left(\frac{r_t{(2\ell+h)}}{r_t{(h)}}\frac{1+\sqrt{f(r_t{(2\ell+h)})}}{1+\sqrt{f(r_t{(h)})}}\right).
\eea
Using eq.\eqref{lengthheed1}, the above expression can be rewritten as follows
\bea
E_W=\frac{c}{6}\log\frac{\tanh\frac{\pi(2\ell+h)T}{2}}{\tanh\frac{\pi hT}{2}}.
\eea
Hence using the above result we can find EWCS in different scaling regimes. For $hT\ll\ell T\ll 1$ one finds
\bea\label{loweopd3}
E_W= \frac{c}{6}\log\frac{2\ell}{h}-\frac{c}{18}(\pi\ell T)^2+\cdots,
\eea
where the first term is just the zero temperature $E_W$. The second term with a negative sign shows that finite temperature reduces $E_W$ and therefore two subsystems becomes more disentangled. On the other hand for $hT\ll 1 \ll \ell T$ we have
\bea\label{higheopd3}
E_W=- \frac{c}{6}\log\frac{\pi hT}{2}+ \frac{c}{6}\log\tanh(\pi \ell T)
%+ \frac{c}{3}\pi hT e^{-2\pi \ell T}
+\cdots,
\eea
where we have neglected terms those suppress exponentially with $\ell T$. Note that the first inequality in eq.\eqref{inequality} satisfied in both low and high temperature limits. Also taking the limit for adjacent subregions $h\rightarrow 0$ in the above result, we see that the EWCS diverges.

\begin{figure}
\begin{center}
\includegraphics[scale=0.6]{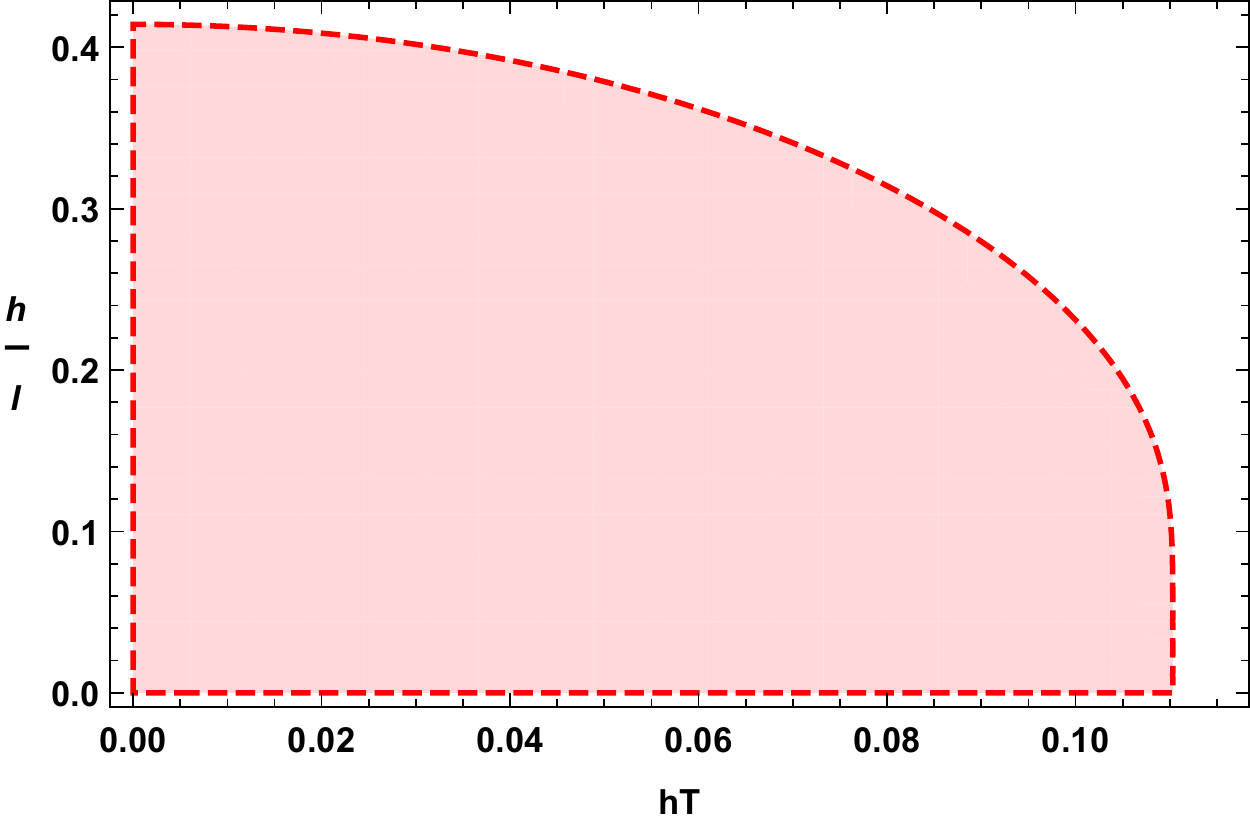}
\hspace*{0.5cm}
\includegraphics[scale=0.85]{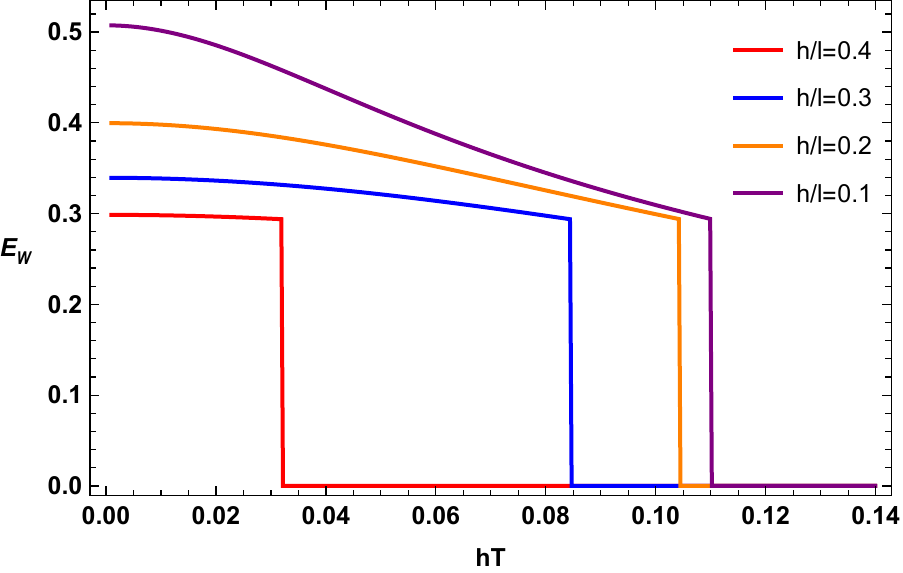}
\end{center}
\caption{\textit{Left}: parameter space for $d=1$ where $E_W$ is nonzero only in the red shaded region. \textit{Right}: $E_W$ as a function of $hT$ for different values of $h/\ell$. In all these cases $E_W$ undergoes a discontinuous phase transition beyond which it is identically zero. Here we set $c=1$.}
\label{fig:d3eop}
\end{figure}

The behavior of $E_W$ in a two dimensional field theory can be read off from Fig.\ref{fig:d3eop}. The left panel shows the two dimensional parameter space restricted by the $E_W\neq0$ condition. Note that it is convenient to define dimensionless variables $hT$ and $\frac{h}{\ell}$. The nonzero $E_W$ corresponds to the red shaded region where the correlation between two subregions are nonvanishing. According to this plot, keeping the length of $A$ and $B$ fixed while their separation increases, the EWCS shows a discontinuous phase transition, such that $E_W=0$ when the two regions are distant enough. Further EWCS is a monotonically decreasing function of temperature such that in high temperature limit vanishes. It is worth to mention that precisely, the similar situation arose in \cite{Fischler:2012ca} where the structure of HMI has been investigated, although the HMI transition is continuous.

\subsubsection{EWCS in $d>1$}
\begin{figure}
\begin{center}
\includegraphics[scale=0.8]{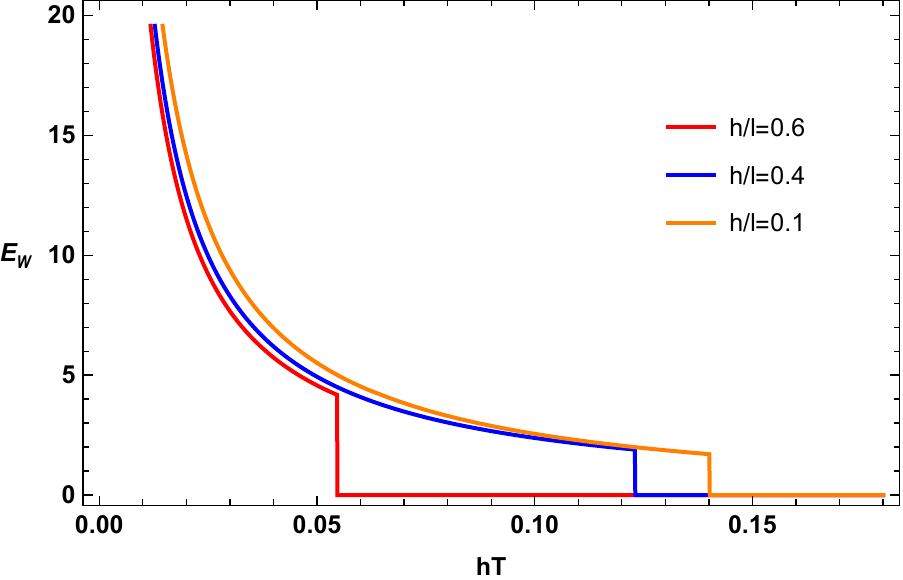}
\hspace*{0.5cm}
\includegraphics[scale=0.84]{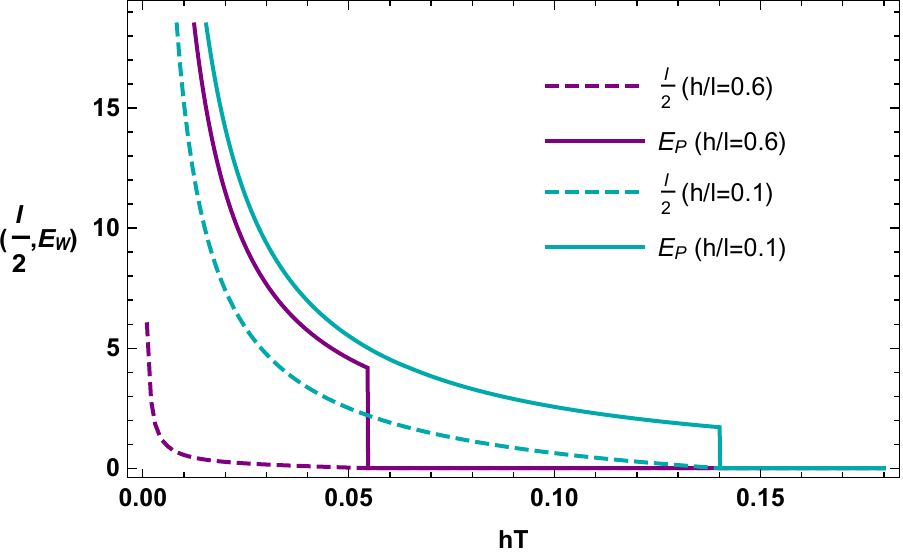}
\end{center}
\caption{\textit{Left}: $E_W$ as a function of $hT$ for different values of $h/\ell$ in $d=2$. In all these cases $E_W$ undergoes a discontinuous phase transition beyond which it is identically zero. \textit{Right}: Checking $I/2\leq E_W$. Here we set $H/G_N=1$.}
\label{fig:d4eop}
\end{figure}

In this case the integral in eq.\eqref{eopfunc} can be rewritten as follows\footnote{Note that although evaluating this integral gives an exact result\cite{Yang:2018gfq}
\bea
E_W=\frac{H^{d-1}}{4(1-d)G_N}\left(\frac{\sqrt{f(r)}}{r^{d-1}}-\frac{d-3}{4}\frac{r^2}{r_0^{d+1}} \,{_2F_1}\left(\frac{1}{2},\frac{2}{d+1},\frac{d+3}{d+1},\frac{r^{d+1}}{r_0^{d+1}}\right)\right)\bigg|_{r_t{h}}^{r_t{(2\ell+h)}}\nonumber,
\eea
using the systematic expansion method is more tractable to find the low and high temperature corrections of EWCS.
}
\bea\label{eopddim}
E_W&=&\frac{H^{d-1}}{4G_N}\int_{r_t{(h)}}^{r_t{(2\ell+h)}}\frac{dr}{r^{d}}\sum_{n=0}^{\infty}\frac{\Gamma\left(n+\frac{1}{2}\right)}{\sqrt{\pi}\Gamma(n+1)}\left(\frac{r}{r_0}\right)^{n(d+1)},\nonumber\\
&=&\frac{H^{d-1}}{4G_N}\sum_{n=0}^{\infty}\frac{\Gamma\left(n+\frac{1}{2}\right)}{\sqrt{\pi}\Gamma(n+1)}\frac{1}{r_0^{n(d+1)}}\frac{r_t(2\ell+h)^{n(d+1)-d+1}-r_t(h)^{n(d+1)-d+1}}{n(d+1)-d+1}.
\eea
Using this expression one can study the behavior of $E_W$ as a function of $h$, $\ell$ and $T$. While various dimensions do not yield
the same results quantitatively, they still agree at a qualitative level. Therefore in the following we focus on $d=2$ case. The left panel of Fig.\ref{fig:d4eop} shows $E_W$ as a function of $hT$ for different values of $\frac{h}{\ell}$. Once again we observe that $E_W$ is a monotonically decreasing function of temperature and vanishes for far away regions. As a consistency check in the right panel we plot both $E_W$ and $\frac{I}{2}$ to see weather eq.\eqref{inequality} satisfied or not. 

It is also interesting to look at low and high temperature expansions of $E_W$. At the low temperature, considering the $h T\ll \ell T\ll 1$ limit and using eq.\eqref{rtsmalltemp} for $r_t(h)$ and $r_t(2\ell+h)$, eq.\eqref{eopddim} yields
\bea
E_W=E_W(T=0)-2\mathcal{C}_0\mathcal{C}_1\frac{H^{d-1}}{G_N}\frac{c'}{c}\ell(\ell+h)T^{d+1}+\cdots,
\eea
where 
\bea
c'=\frac{1}{d}\left(1-\frac{\sqrt{\pi}d(d+2)}{2^{1/d}}\frac{\Gamma\left(\frac{d+2}{2d}\right)}{\Gamma\left(\frac{1}{2d}\right)^2}\right),
\eea
and $E_W({T=0})$ is the EWCS at zero temperature, which in this case is given by 
\bea\label{eopt0}
E_W(T=0)=\frac{\mathcal{C}_0}{c}\frac{H^{d-1}}{4G_N}\left(-\frac{1}{h^{d-1}}+\frac{1}{(2\ell+h)^{d-1}}\right).
\eea
On the other hand in $h T\ll 1\ll \ell T$ limit corresponding to $r_t(h)\ll r_0$ and $r_t(2\ell+h)\rightarrow r_0$ we reexpress eq.\eqref{eopddim} as follows 
\bea\label{eopddimseries}
E_W=\frac{H^{d-1}}{4G_N}\sum_{n=0}^{\infty}\frac{1}{n(d+1)-d+1}\frac{\Gamma\left(n+\frac{1}{2}\right)}{\sqrt{\pi}\Gamma(n+1)}\left(\frac{r_t(2\ell+h)^{n(d+1)-d+1}}{r_0^{n(d+1)}}-\frac{r_t(h)^{n(d+1)-d+1}}{r_0^{n(d+1)}}\right).
\eea
It is easy to see that in the large $n$ limit the first infinite series behaves as $\frac{1}{n^{3/2}}\left(\frac{r_t(2\ell+h)}{r_0}\right)^{n(d+1)}$, so we can safely consider $r_t(2\ell+h)\rightarrow r_0$ limit. Also to consider $r_t(h)\ll r_0$ limit we keep only the leading order terms in the second infinite series, which yields
\bea\label{eopddimseries2}
E_W=\frac{H^{d-1}T^{d-1}}{4G_N}\left(-\frac{\mathcal{C}_0}{c(hT)^{d-1}}+\left(\frac{4\pi}{d+1}\right)^{d}\mathcal{C}_2'+2\frac{c'}{c}\mathcal{C}_0\mathcal{C}_1(hT)^2\right),
\eea
where $\mathcal{C}_2'=\frac{\Gamma\left(\frac{1-d}{1+d}\right)}{4\sqrt{\pi}\Gamma\left(\frac{3-d}{2(1+d)}\right)}$. The first term in the above expression diverges in $h\rightarrow 0$ limit where the subregions coincide. Further, the second term which is proportional to the area of the entangling region shows that the EWCS obeys an area law even in finite temperature. As we mentioned before, for a mixed thermal state HEE measures both classical and quantum correlations and as a result scales with the volume. Therefore we may conclude that $E_W$ carries more relevant content than HEE as far as computing quantum entanglement is concerned.

\section{EWCS in Theories with Lifshitz Scaling and Hyperscaling Violation}\label{sec:nonrel}
In this section we study the finite temperature contribution to the entanglement wedge cross section in holographic theories with general dynamical critical exponent $z$ and hyperscaling violation exponent $\theta$. These theories admit a fixed point where the system is invariant under the following anisotropic scaling transformation
\begin{equation}
r \to \lambda r, \quad t \to \lambda^z t, \quad \vec{x} \to \lambda \vec{x}, \quad ds \to \lambda^{\frac{\theta}{d}} ds.
\end{equation} 
Various holographic aspects of these theories have been studied in \cite{Dong:2012se,Alishahiha:2012qu,Gath:2012pg,Alishahiha:2014cwa,Tanhayi:2015cax,Alishahiha:2018tep}. In particular, authors of \cite{Fischler:2012uv} have studied the HEE and HMI of these theories at finite temperature.\footnote{It is worth to mention that various aspects of entanglement measures in QFTs with Lifshitz scaling symmetry are studied in \cite{MohammadiMozaffar:2017nri,He:2017wla,MohammadiMozaffar:2017chk,MohammadiMozaffar:2018vmk}. } In what follows, similar to section \ref{sec:finiteT}, we consider an entangling region in the shape of a strip and calculate finite temperature corrections to EWCS. For completeness, we also briefly review the main result of \cite{Fischler:2012uv} about HEE and HMI in theories with  Lifshitz scaling and hyperscaling violation at the finite temperature in the appendix \ref{LHTNR}. 

Let us consider a $(d+2)$-dimensional black brane solution in the Einstein theory of gravity with appropriate matter field (\emph{e.g.} see \cite{Dong:2012se})
\bea \label{metricNR}
ds^2=\frac{1}{r_f^{2\,\theta/d}}\frac{1}{r^{2\frac{d-\theta}{d}}}\left(-\frac{f(r)}{r^{2(z-1)}}dt^2+\frac{dr^2}{f(r)}+d\vec{x}^2\right), \quad f(r)=1-\qty(\frac{r}{r_0})^{d-\theta+z},
\eea
where $r_0$ is the horizon radius. In addition, $r_f$ is a length scale which fixes the dimensions when $\theta\ne 0$ and in the following without loss of generality we set $r_f=1$. As mentioned in \cite{Dong:2012se}, the null energy condition implies some bounds on values of $\theta$ and $z$
\begin{equation}\label{NEC}
(d-\theta)(d(z-1)-\theta)\ge 0, \quad (z-1)(d-\theta+z) \ge 0.
\end{equation}
The temperature and thermal entropy density for eq.\eqref{metricNR} are given by
\bea\label{tempNR}
T=\frac{\vert d-\theta+z\vert}{4\pi r_0^z},\;\;\;\;\;\;\; s_{\rm th}=\frac{1}{4G_N\,}\frac{1}{r_0^{d-\theta}}.
\eea
It is easy to see that the entropy density scales as $s_{\rm th}\sim T^{\frac{d-\theta}{z}}$. For $z=1$, it shows that hyperscaling violating exponent effectively reduces the dimensionality of the model. As we will see in what follows it is a typical role of $\theta$. The thermodynamic stability of black brane solution requires a positive specific heat which restricts $\theta$ and $z$ further
\begin{equation}\label{positiveSH}
\frac{d-\theta}{z}\ge 0.
\end{equation}
In addition, as argued in \cite{Dong:2012se}, entanglement entropy analysis at zero temperature as well as string theory realization of hyperscaling violating geometries imply inconsistency when $\theta>d$. So, in what follows we assume $d-\theta>0$ and $z\ge1$ which satisfy both eqs. \eqref{positiveSH} and \eqref{NEC}.

\subsection{Low and High Temperature Behavior of EWCS}\label{EoPTLH}
Now we calculate $E_W$ for holographic theories with hyperscaling violating geometry, using the holographic prescription \cite{Takayanagi:2017knl,Nguyen:2017yqw}. 
The $E_W$ for configuration depicted in Fig.\ref{fig:regionseop} for a QFT dual to eq.\eqref{metricNR} can be written as 
\bea\label{eopfuncNR}
E_W=\frac{H^{d-1}}{4G_N }\int_{r_t{(h)}}^{r_t{(2\ell+h)}}\frac{dr}{r^{\dt}\sqrt{1-\frac{r^{\dt+z}}{r_0^{\dt+z}}}},
\eea
where an effective dimension $\dt=d-\theta$ is defined. It is easy to see that, for $\theta=d-1$ and $z=1$, this expression reduces to what we  have obtained for a three dimensional theory in section \ref{sec:threedim}. Moreover, for $\theta=d$, as eqs.\eqref{eomtempNR} and \eqref{heetempNR} show, the RT surface lies on boundary slice $r=\epsilon$ and there is no turning point at all. So, the notion of entangling wedge is not well defined. Indeed, in this case the HEE exhibits an extensive violation of area law \cite{Dong:2012se}. In the rest of this section we neglect these two special cases and calculate the EWCS at low and high temperature limits.

\begin{figure}
	\begin{center}
		\includegraphics[scale=0.8]{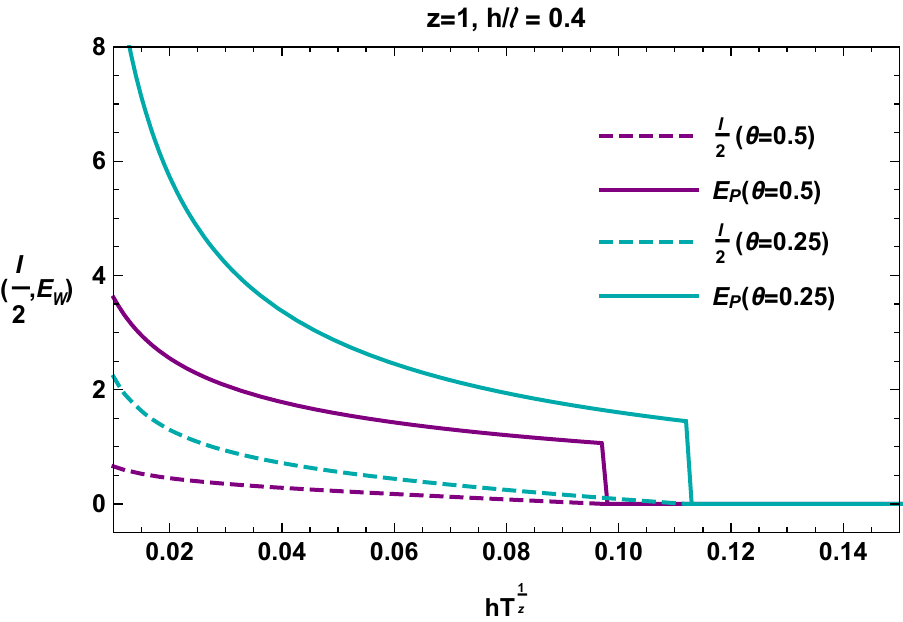}
		\hspace*{0.5cm}
		\includegraphics[scale=0.8]{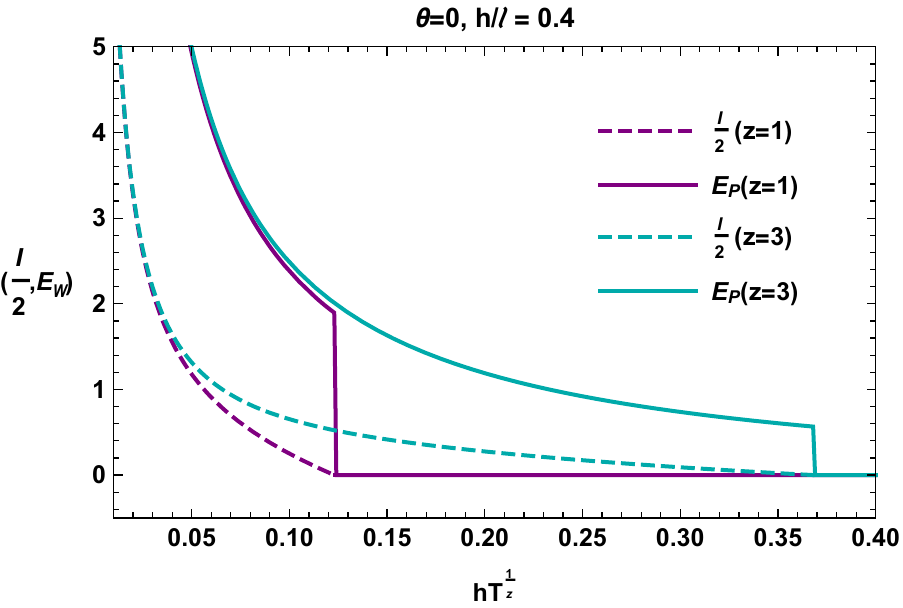}
	\end{center}
	\caption{EWCS as a function of $hT^{\frac{1}{z}}$ for different values of $z$ and $ \theta$. We also plot $\frac{I}{2}$ to check eq.\eqref{inequality}. Here we set $H/G_N=1$ and $d=2$.  }
	\label{fig:eopHMINR}
\end{figure}

\begin{figure}
	\begin{center}
		\includegraphics[scale=0.8]{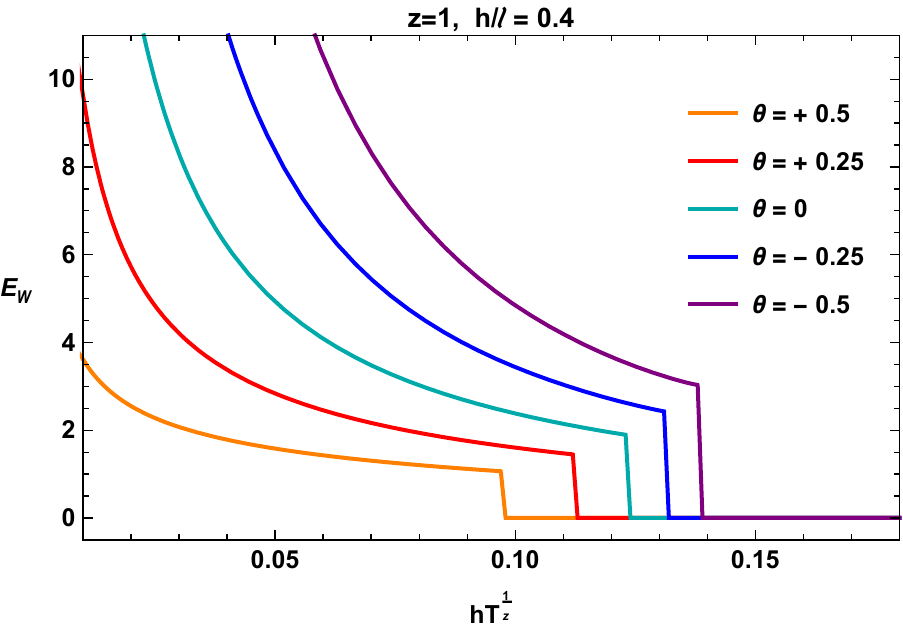}
		\hspace*{0.5cm}
		\includegraphics[scale=0.8]{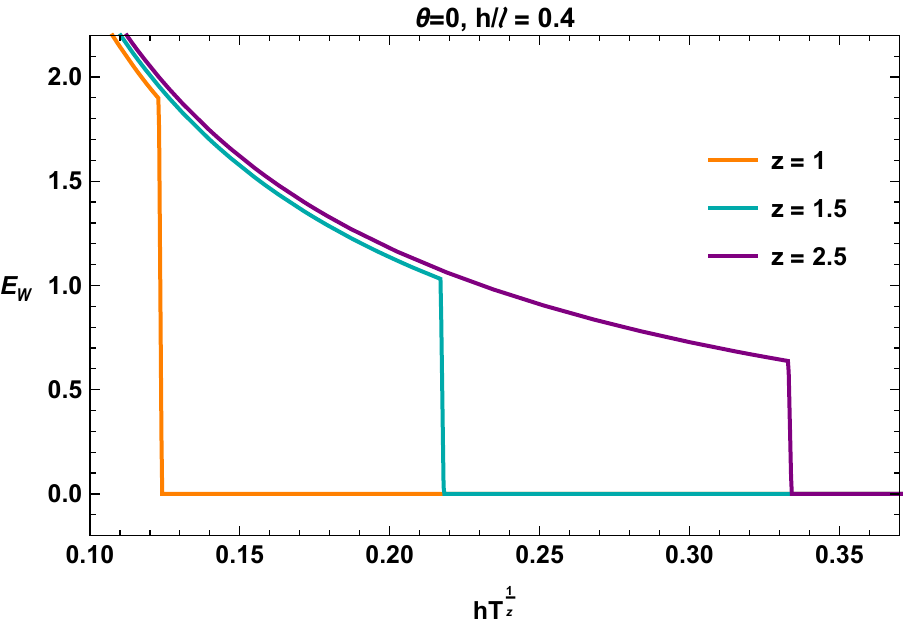}
	\end{center}
	\caption{EWCS as a function of $hT^{\frac{1}{z}}$ for different values of $\theta$ and $z$. \textit{Left}: $E_W$ decreases as $\theta$ increases.  \textit{Right}: $E_W$ increases as $z$  increases. Here we set $H/G_N=1$ and $d=2$.  }
		\label{fig:eopNR}
\end{figure}

%\subsubsection{EoP in $\theta\neq d$}
Employing binomial series for the integrand of eq.\eqref{eopfunc}, the EWCS integral can be rewritten as follows
\bea\label{eopddimNR}
E_W&=&\frac{H^{d-1}}{4G_N }\int_{r_t{(h)}}^{r_t{(2\ell+h)}}\frac{dr}{r^{\dt}}\sum_{n=0}^{\infty}\frac{\Gamma\left(n+\frac{1}{2}\right)}{\sqrt{\pi}\Gamma(n+1)}\left(\frac{r}{r_0}\right)^{n(\dt+z)},\nonumber\\
&=&\frac{H^{d-1}}{4G_N}\sum_{n=0}^{\infty}\frac{\Gamma\left(n+\frac{1}{2}\right)}{\sqrt{\pi}\Gamma(n+1)}\frac{1}{r_0^{n(\dt+z)}}\frac{r_t(2\ell+h)^{n(\dt+z)-\dt+1}-r_t(h)^{n(\dt+z)-\dt+1}}{n(\dt+z)-\dt+1}.
\eea
Let us consider $h T\ll \ell T\ll 1$ limit where we can use eq. \eqref{rtsmalltempNR} for $r_t(h)$ and $r_t(2\ell+h)$ to obtain $E_W$ at low temperature limit 
\bea\label{epnonrelT}
E_W=E_W(T=0)-\Ct_0\Ct_1\frac{H^{d-1}}{2G_N}\frac{\ct'}{\ct}\qty((2\ell+h)^{z+1}-h^{z+1})T^{\frac{\dt+z}{z}}+\cdots,
\eea
where $\Ct_0$ and $\Ct_1$ are defined in eq.\eqref{C0C1} and
\bea
\ct'=\frac{1}{2}\left(\frac{1+z}{\dt}-\frac{2 \Gamma\qty(\frac{\dt+1}{2\dt})\Gamma\qty(\frac{3\dt+z+1}{2\dt})}{\Gamma\qty(\frac{1}{2\dt})\Gamma\qty(\frac{2\dt+z+1}{2\dt})}\right).
\eea
In addition, $E_W({T=0})$ corresponds to EWCS at zero temperature
\bea \label{epNR_zero_temp}
E_W(T=0)=\frac{\Ct_0}{\ct}\frac{H^{d-1}}{4G_N\,\rft}\left(-\frac{1}{h^{\dt-1}}+\frac{1}{(2\ell+h)^{\dt-1}}\right).
\eea
Considering $z>1$ and $\dt>0$ case, eq.\eqref{epnonrelT} shows that $E_W$ decreases with temperature, as expected. In addition, eq.\eqref{epNR_zero_temp} shows that with decreasing effective dimension $\dt$ ($\theta$ increasing), $E_W$ decreases. 
On the other hand in $h T\ll 1\ll \ell T$ limit corresponding to $r_t(h)\ll r_0$ and $r_t(2\ell+h)\rightarrow r_0$ we reexpress eq.\eqref{eopddimNR} as follows 
\bea\label{eopddimseriesNR}
E_W=\frac{H^{d-1}}{4G_N\,\rft}\sum_{n=0}^{\infty}\frac{1}{n(\dt+z)-\dt+1}\frac{\Gamma\left(n+\frac{1}{2}\right)}{\sqrt{\pi}\Gamma(n+1)}\Bigg(\frac{r_t(2\ell+h)^{n(\dt+z)-\dt+1}}{r_0^{n(\dt+z)}}-\frac{r_t(h)^{n(\dt+z)-\dt+1}}{r_0^{n(\dt+z)}}\Bigg{)}.
\eea
%\begin{align}\label{eopddimseriesNR}
%E_P&=\frac{H^{d-1}}{4G_N\,\rft}\sum_{n=0}^{\infty}\frac{1}{n(\dt+z)-\dt+1}\frac{\Gamma\left(n+\frac{1}{2}\right)}{\sqrt{\pi}\Gamma(n+1)}\nn\\
%&\times \Bigg(\frac{r_t(2\ell+h)^{n(\dt+z)-\dt+1}}{r_0^{n(\dt+z)}}-\frac{r_t(h)^{n(\dt+z)-\dt+1}}{r_0^{n(\dt+z)}}\Bigg{)}.
%\end{align}
In the large $n$ limit the first infinite series behaves as $\frac{1}{n^{3/2}}\left(\frac{r_t(2\ell+h)}{r_0}\right)^{n(\dt+z)}$, so we can safely consider $r_t(2\ell+h)\rightarrow r_0$ limit. Also to consider $r_t(h)\ll r_0$ limit, we keep only the first two terms in the second infinite series, which yields
	\bea\label{eopddimseries2NR}
	E_W=\frac{H^{d-1}T^\frac{{\dt-1}}{z}}{4G_N\,\rft}\left(-\frac{\Ct_0}{\ct(hT^{\frac{1}{z}})^{\dt-1}}+\left(\frac{4\pi}{\vert\dt+z\vert}\right)^{\frac{\dt}{z}}\Ct_2'+2\frac{\ct'}{\ct}\Ct_0\Ct_1(hT^\frac{1}{z})^{z+1}\right),
	\eea
	where
	 \be
	\Ct_2'=\qty(\frac{\vert\dt+z\vert}{4\pi})^{\frac{1}{z}}\sum_{n=0}^{\infty}\frac{1}{n(\dt+z)-\dt+1}\frac{\Gamma\left(n+\frac{1}{2}\right)}{\sqrt{\pi}\Gamma(n+1)}=\frac{1}{1-\dt}\qty(\frac{\vert\dt+z\vert}{4\pi})^{\frac{1}{z}}{_2F_1}\left(\frac{1}{2},\frac{1-\dt}{\dt+z},\frac{z+1}{\dt+z},1\right).
	\ee
In Figs.\ref{fig:eopHMINR} and \ref{fig:eopNR} we plot $E_W$ as a function of $hT^{\frac{1}{z}}$ for different values of $z$ and $\theta$. Once again we observe that $E_W$ is a monotonically decreasing function of temperature and vanishes for distant enough regions. 
As illustrated in these figures, the transition point depends on the value of dynamical and hyperscaling violation exponents. In particular, in Fig.\ref{fig:eopHMINR} we have compared EWCS and HMI for different values of $z$ and $\theta$ which in all cases eq.\eqref{inequality} satisfied. Based on these figures $E_W$ is an increasing function of the dynamical exponent, i.e., $z$, and the transition which happens in large separation or high temperature limit is slower in comparing to the relativistic case with $z=1$. Such a behavior is not surprising because as discussed in \cite{MohammadiMozaffar:2017nri} the spatial correlations between subregions become stronger for larger values of $z$. On the other hand as is clear from the graphs, $E_W$ is monotonically decreasing as the hyperscaling violating exponent increases. As we mentioned before hyperscaling violation leads to an effective dimension $\tilde{d}=d-\theta$ and therefore in a theory with nonvanishing $\theta$, effective spatial dimension decreases (increases) for larger (smaller) values of hyperscaling violating exponent . Hence for larger values of $\theta$ we expect that the spatial correlations between subregions decrease and the resultant $E_W$ decreases.\footnote{We would like to thank Ali Mollabashi for useful comments about this point.}

\section{Corner Contributions to EWCS}\label{sec:eopcorner}

\begin{figure}
\begin{center}
\includegraphics[scale=1]{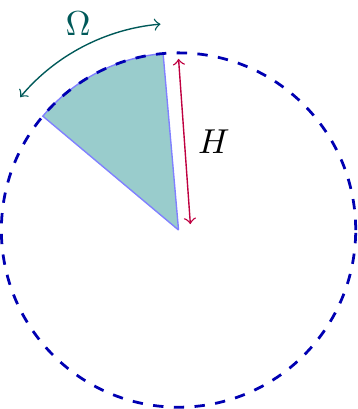}
\hspace{2cm}
\includegraphics[scale=1]{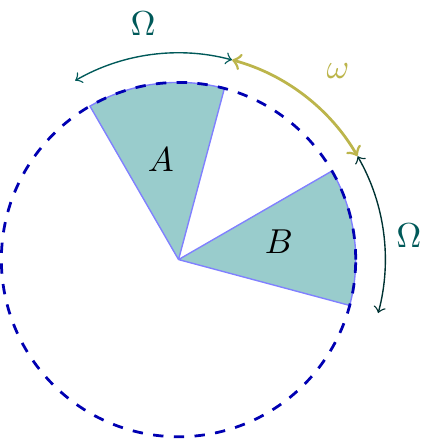}
\end{center}
\caption{Schematic configuration for computing HEE (left) and EWCS (right) in $d=2$ when the entangling region contains a corner.}
\label{fig:corner}
\end{figure}

In this section we study the corner contribution to the EWCS for holographic CFTs dual to Einstein gravity. In the holographic context, considering singular entangling surfaces was first done in \cite{Hirata:2006jx}. Various features of holographic entanglement entropy for regions with a singular boundary such as cone and crease have been studied, e.g., see \cite{Myers:2012vs,Bueno:2015rda,Alishahiha:2015goa,Seminara:2017hhh}. A key feature of these studies is the appearance of a new logarithmic term in the HEE which depends on the central charge of the underlying CFT. The coefficient of this universal term depends on the opening angle of the corresponding singular surface such that in the smooth limit where the singularity disappears,  vanishes. It is worthwhile to point that, corner contributions to other entanglement/information measures is also studied , e.g., see \cite{Mozaffar:2015xue,Bakhshaei:2017qud}. In particular, for a specific configuration (see Fig.\ref{fig:corner}) the HMI becomes UV divergent when the singular subregions coincide\cite{Mozaffar:2015xue}. In the following considering the same setup, we would like to investigate to what extent these singularities in the boundary of entangling regions modify the behavior of EWCS. To begin with, we will compute EWCS for a union of kinks in $d=2$ in the next subsection.

\subsection{EWCS for a Union of Kinks in AdS$_4$}

In this section, in order to find the EWCS for a union of kinks, we use the following 4-dimensional bulk geometry
\bea\label{metric22}
ds^2=\frac{1}{r^2}\left(-dt^2+dr^2+d\rho^2+\rho^2 d\phi^2\right),
\eea
where we have written the boundary spatial directions in polar coordinates. A kink entangling region is specified by
\bea\label{bdy}
t={\rm const.},\;\;\;\;0\leq \rho \leq H,\;\;\;-\Omega\leq 2\phi \leq \Omega,
\eea
where $H$ is the IR cut-off (see Fig.\ref{fig:corner}). Due to the scaling symmetry of the AdS background and the fact that $\rho$ is the only scale in our setup, for the RT surface we choose the parametrization $r(\rho, \phi)=\rho \,\Phi(\phi)$.
% such that $\Phi(\pm \Omega)=0$. 
 In this case the corresponding HEE functional becomes
\bea\label{heefunccone}
S=\frac{1}{4G_N}\int_{\frac{\epsilon}{\Phi_t}}^{H}\frac{d\rho}{\rho}\int_0^{\frac{\Omega}{2}-\delta}\frac{d\phi}{\Phi}\sqrt{1+\Phi^2+\Phi'^2},
\eea
where $\Phi_t\equiv\Phi(0)$ is the turning point given by $\Phi'(0)=0$ and $\epsilon=\rho\, \Phi(\frac{\Omega}{2}-\delta)$ is the UV cut off. 
Since there is no explicit $\phi$ dependence, the corresponding Hamiltonian is a conserved quantity. Therefore, we find the following first integral
\bea\label{hamil}
\frac{1+\Phi^2}{\Phi^2\sqrt{1+\Phi^2+\Phi'^2}}=\frac{\sqrt{1+\Phi_t^2}}{\Phi_t^2}.
\eea
Using the above equation and the boundary condition eq.\eqref{bdy}, the opening angle is
\bea\label{openingangle}
\Omega=\int_0^{\Phi_t}d\Phi\frac{\Phi^2\sqrt{1+\Phi_t^2}}{\sqrt{1+\Phi^2}\sqrt{\Phi_t^4(1+\Phi^2)-\Phi^4(1+\Phi_t^2)}}.
\eea
Substituting eq.\eqref{hamil} back into the expression for the HEE eq.\eqref{heefunccone}, we finally obtain
\bea\label{heekink}
S(\Omega)=\frac{1}{2G_N}\frac{H}{\epsilon}-s(\Omega)\log\frac{H}{\epsilon}-\frac{\pi}{4G_N\Phi_t}-s(\Omega)\log\Phi_t+\mathcal{O}\left(\frac{\epsilon}{H}\right),
\eea
where
\bea
s(\Omega)=\frac{1}{2G_N}\int_0^{\infty}du\left(1-\frac{\sqrt{1+\Phi_t^2(1+u^2)}}{\sqrt{2+\Phi_t^2(1+u^2)}}\right).
\eea
The precise expression for the coefficient of the new universal term can be obtained in certain limits. In particular in $\Omega \rightarrow 0$ limit where we have a sharp corner, one finds\cite{Hirata:2006jx,Bueno:2015rda}
%\bea
%s(\Omega \rightarrow 0)\sim \frac{\kappa}{\Omega}+\cdots,\;\;\;\;\kappa=\frac{1}{2\pi G_N}\Gamma\left(\frac{3}{4}\right)^4,
%\eea
\bea\label{sharpcorner}
s(\Omega \rightarrow 0)\sim \frac{\kappa}{\Omega}+\cdots,\;\;\;\;\;\;\;\kappa=\frac{\pi^2}{6}\Gamma\left(\frac{3}{4}\right)^4 C_T,
\eea
where $C_T\equiv\frac{3}{\pi^3G_N}$ is the central charge appearing in two-point function of the stress tensor for the underlying CFT.\footnote{The explicit expression for the corresponding two-point function is 
\bea
\langle T_{\mu\nu}(r)T_{\alpha\beta}(0)\rangle=\frac{C_T}{r^{2d}}\mathcal{I}_{\mu\nu,\alpha\beta}(r),
\eea
where $\mathcal{I}_{\mu\nu,\alpha\beta}$ is a tensor fixed by symmetry.}
These results are easily extended to general multipartite subregions, to compute other entanglement measures, e.g., mutual and tripartite information (see \cite{Mozaffar:2015xue} for a complete discussion).

In order to compute $E_W$ we consider the configuration depicted in the right panel of Fig.\ref{fig:corner}. Once again we focus our attention on the connected configuration for RT surfaces where both the HMI and EWCS are nonzero. Note that due to the axial symmetry we expect that the minimal cross section of entanglement wedge locates at $\phi=0$. Using eq.\eqref{metric22}, the EWCS functional is given by
\bea
E_{W}=\frac{1}{4G_N}\int_{\epsilon}^{H}d\rho\int_{\rho \Phi_t({\omega})}^{\rho \Phi_t({2\Omega+\omega})}\frac{dr}{r^2},
\eea
where $\Phi_t({\omega})$ and $\Phi_t({2\Omega+\omega})$ denote the turning points of the corresponding minimal surfaces. The above integral can be evaluated explicitly yielding
\bea
E_{W}=\frac{1}{4G_N}\left(\frac{1}{\Phi_t({\omega})}-\frac{1}{\Phi_t({2\Omega+\omega})}\right)\log\frac{H}{\epsilon},
\eea
which demonstrates that EWCS is divergent when the subregions coincide. It is worth to mention that precisely, the similar situation arose in \cite{Mozaffar:2015xue} in investigating the structure of HMI, although the transition is continuous. Considering the case where we have two adjacent sharp corners, i.e., $\omega\ll \Omega\ll 1$ and using eq.\eqref{sharpcorner} we may further simplify the result to
\bea\label{eopcorner}
E_{W}=\frac{\sqrt{\pi/2}}{\Gamma^2(3/4)}\kappa\left(\frac{1}{\omega}-\frac{1}{\Omega}\right)\log\frac{H}{\epsilon}\sim\frac{\sqrt{\pi/2}}{\Gamma^2(3/4)}\frac{\kappa}{\omega}\log\frac{H}{\epsilon}.
\eea
It is important to mention that the above result can be obtained using a conformal map relating the corner geometry to a strip in four dimensions. A similar
derivation to the one presented for HMI in \cite{Mozaffar:2015xue} holds in the present case which shows that the above expression reduces to eq.\eqref{eopt0} for $d=2$.
As another consistency check, we note that the EWCS should satisfy eq.\eqref{inequality}. Indeed, as shown in \cite{Mozaffar:2015xue} the HMI in this particular limit is given by
\bea
I=\kappa\left(\frac{1}{\omega}-\frac{2}{\Omega}+\frac{1}{2\Omega+\omega}\right) \log\frac{H}{\epsilon}\sim\frac{\kappa}{\omega} \log\frac{H}{\epsilon}.
\eea
In comparing the above expression with eq.\eqref{eopcorner}, we see that the constraint on the lower bound of EWCS satisfied.

%{\color{red}
%\textbf{(ii) smooth limit $\omega\sim 0, \;\Omega\sim \pi$}
%\bea
%E_{P}=\frac{L^2}{4G_N}\left(\frac{2\sqrt{\pi}\Gamma(3/4)}{\Gamma(1/4)\omega}+\Omega-1\right)\log\frac{H}{\tilde{\epsilon}}\sim\frac{\sqrt{\pi/2}}{\Gamma^2(3/4)}\frac{\kappa}{\omega}\log\frac{H}{\tilde{\epsilon}},
%\eea
%the HMI
%\bea
%\frac{1}{2}I\sim \frac{\kappa}{\omega} \log\frac{H}{\epsilon},
%\eea
%}

\subsection{EWCS for a Union of Creases in AdS$_{d+2}$}

In this section we will compute the EWCS in the presence of singular regions in higher dimensions. The calculations are analogous to those for three dimensions. Consider the following bulk geometry
\bea\label{metricdd}
ds^2=\frac{1}{r^2}\left(-dt^2+dr^2+d\rho^2+\rho^2 d\phi^2+\sum_{i=1}^{d-2}dx_i^2\right).
\eea
In this case the entangling region is specified by
\bea\label{bdyd}
t={\rm const.},\;\;\;\;0\leq \rho \leq H,\;\;\;-\Omega\leq 2\phi \leq \Omega,\;\;\;0<x_i<\tilde{H},
\eea
where $H$ and $\tilde{H}$ are the IR regulators where in the following we set $\tilde{H}=H$. Using the scaling symmetry of the background and assuming $r(\rho,\phi)=\rho\, \Phi(\phi)$, the corresponding HEE functional becomes
\bea
S=\frac{H^{d-2}}{4G_N}\int_{\frac{\epsilon}{\Phi_t}}^{H}\frac{d\rho}{\rho^{d-1}}\int_{0}^{\frac{\Omega}{2}-\delta}d\phi\frac{\sqrt{1+\Phi^2+\Phi'^2}}{\Phi^{d}}.
\eea
Once again, since there is no explicit $\phi$ dependence, we have a first integral\cite{Myers:2012vs}
\bea
\mathcal{K}_d\equiv\frac{(1+\Phi^2)^{\frac{d}{2}}}{\Phi^d\sqrt{1+\Phi^2+\Phi'^2}}=\frac{(1+\Phi_t^2)^{\frac{d-1}{2}}}{\Phi_t^d}.
\eea
This eventually leads to the following expression for the opening angle and HEE
\bea\label{heedcrease}
\Omega=2\mathcal{K}_d\int_0^{\Phi_t}d\Phi\frac{\Phi^d}{\sqrt{1+\Phi^2}\sqrt{\left(1+\Phi^2\right)^{d-1}-\mathcal{K}_d^2\Phi^{2d}}},\nonumber\\
S=\frac{H^{d-2}}{2G_N}\left(\frac{H}{(d-1)\epsilon^{d-1}}+\frac{\mathcal{F}(\Omega)}{(d-2)\epsilon^{d-2}}\right)+\mathcal{O}(\epsilon),
\eea
where we have defined
\bea
\mathcal{F}(\Omega)=-\frac{1}{\Phi_t(\Omega)}-\int_{0}^{\Phi_t(\Omega)}\frac{d\Phi}{\Phi^2}\left(1+\frac{\sqrt{1+\Phi^2+\Phi'^2}}{\Phi'}\right).
\eea
It is worth to mention that the second divergent term in eq.\eqref{heedcrease} is produced by the singularity in the entangling surface and vanish
when the surface is smooth. We would like to stress that this contribution is due to adding a flat locus to the kink\cite{Myers:2012vs}. Further in $d=2$ this term modified and we recover a universal logarithmic contribution.  

Turning to EWCS, we expect that the minimal cross section of entanglement wedge locates at $\phi=0$. Using eq.\eqref{metricdd}, one finds
\bea
E_{W}=\frac{H^{d-2}}{4G_N}\int_{\epsilon}^{H}d\rho\int_{\rho \Phi_t({\omega})}^{\rho \Phi_t({2\Omega+\omega})}\frac{dr}{r^d}=\frac{H^{d-2}}{4(d-1)G_N}\left(\frac{1}{\Phi_t({\omega})}-\frac{1}{\Phi_t({2\Omega+\omega})}\right)\int_{\epsilon}^{H}\frac{d\rho}{\rho^{d-1}}.
\eea
Evaluating the above integral, we are left with
\bea
E_{W}=\frac{1}{4(d-1)(d-2)G_N}\left(\frac{1}{\Phi_t({\omega})}-\frac{1}{\Phi_t({2\Omega+\omega})}\right)\frac{H^{d-2}}{\epsilon^{d-2}},
\eea
%\bea
%E_{P}=\frac{1}{4(d-1)(d-2)G_N}\left(\frac{1}{\Phi_t({\omega})}-\frac{1}{\Phi_t({2\Omega+\omega})}\right)\left(\frac{H^{d-2}}{\epsilon^{d-2}}-1\right),
%\eea
which is divergent and obeys the area law. 
\begin{figure}
\begin{center}
\includegraphics[scale=1]{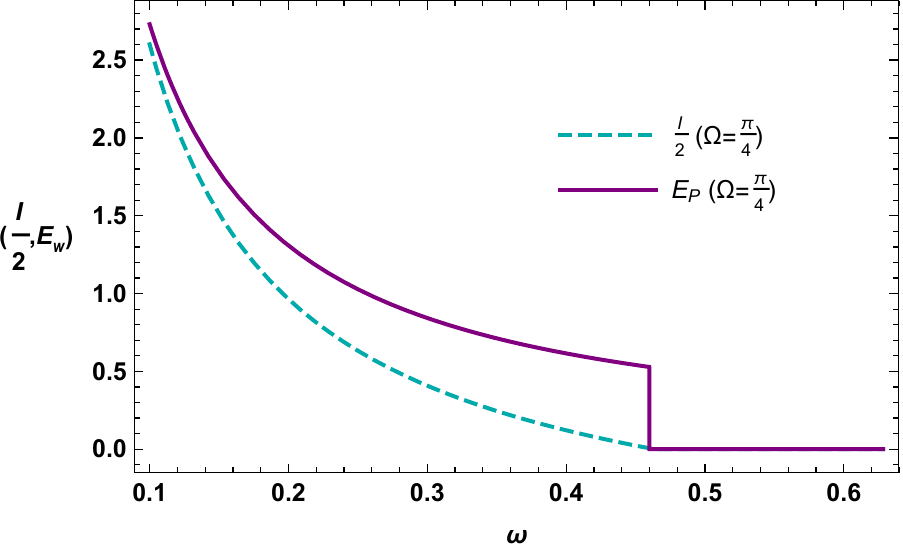}
\end{center}
\caption{EWCS (HMI) as a function of $\omega$ in $d=3$ that undergoes a discontinuous (continuous) phase transition beyond which it is identically zero. Here we normalize these quantites by a factor of $2G_N\epsilon/H$.}
\label{fig:eopcreasemath}
\end{figure}
The final result for $E_W$ in various dimensions agree at a qualitative level, so we just consider $d=3$ case. In Fig.\ref{fig:eopcreasemath} we demonstrate the EWCS as a function of $\omega$ for $\Omega=\frac{\pi}{4}$. We observe that $E_W$ is a monotonically decreasing function of the angular separation between the two subregions and for distant enough regions $E_W= 0$. Further, as a consistency check in this figure we also plot $\frac{I}{2}$ to see weather eq.\eqref{inequality} satisfied or not. Here, it is worth mentioning that, it was shown in \cite{Mozaffar:2015xue} that HMI for a union of creases in $\omega \ll \Omega$ limit is given by the following expression\footnote{Note that in $\Omega \ll \omega$ limit the HMI vanishes.}
\bea
I=\frac{1}{2(d-2)G_N}\left(2\mathcal{F}(\Omega)-\mathcal{F}(2\Omega+\omega)-\mathcal{F}(\omega)\right)\frac{{H}^{d-2}}{\epsilon^{d-2}},
\eea
which is divergent.

\section{Conclusions and Disscusions}\label{sec:conclusions}

In this paper, we explored the general behavior of entanglement wedge cross section (EWCS) in various geometries and for different entangling regions. Based on different holographic interpretation of $E_W$, this quantity may be dual to entanglement of purification, logarithmic negativity and reflected entropy\cite{Takayanagi:2017knl,Nguyen:2017yqw,Kudler-Flam:2018qjo,Dutta:2019gen}. In the following we would like to first summarize our main results and then continue with discussing some further problems. 

\begin{itemize}
\item In a two dimensional relativistic quantum field theory $E_W$ is a monotonically decreasing function of temperature such that at higher temperature the two subsystems becomes more disentangled. Also keeping the length and separation of the subregions fixed while temperature increases, the $E_W$ shows a discontinuous phase transition, such that $E_W=0$ when $T$ is high enough. In the bulk, the vanishing of $E_W$ results because of the disconnected configuration for the RT surfaces and the fact that in this case the corresponding entanglement wedge is disconnected, e.g., see Fig.\ref{fig:eopregions}. 

\item In higher dimensions considering a strip entangling region, the qualitative behaviors of $E_W$ at finite temperature are very similar to $d=1$ case. A key observation is that the $E_W$ obeys an area law even in finite temperature where the HEE shows a volume law. Therefore one may regard the $E_W$ as a more appropriate measure of quantum correlations for thermal mixed states.\footnote{It is important to mention that regarding other basic properties of correlation measures for
mixed states and the holographic interpretation of $E_W$ in terms of EoP, the EoP fails to be an ideal measure of entanglement, because it is not monotonically decreasing under LOCC. }

\item In a nonrelativistic field theory with nontrivial dynamical and hyperscaling exponents, $E_W$ is a monotonically decreasing function of temperature and the separation between subregions. In this case the transition point after that $E_W$ vanishes, depends on the value of $z$ and $\theta$. In particular $E_W$ is an increasing function of the dynamical exponent and the transition which happens in large separation or high temperature limit is slower in comparing to the relativistic case with $z=1$. As we mentioned, the physical reason behind this is that the quantum correlations between subregions increase as one increases $z$. On the other hand, $E_W$ is monotonically decreasing as the hyperscaling violating exponent increases. In a field theory with nonvanishing $\theta$, effective spatial dimension,i.e, $\tilde{d}=d-\theta$, decreases for larger values of hyperscaling violating exponent and therefore for larger values of $\theta$ we expect that the spatial correlations between subregions decrease and the resultant $E_W$ decreases. 

\item Considering an entangling region with singular boundary, we demonstrated that $E_W$ is divergent when the subregions coincide. In particular for a three dimensional boundary theory we found a universal contribution to $E_W$ due to the presence of corner where the coefficient is proportional to the central charge for the underlying CFT. Moreover, considering a singular region in higher dimensions we verified that the corresponding $E_W$ obeys area law. 
\end{itemize}

We can extend this study to different interesting directions. A key feature of $E_W$ is the discontinuous phase transition which happens at large separation or high temperature. As we mentioned before, the corresponding (continuous) phase transition of HMI is a reminiscent of the large $N$ limit of the dual field theory and it disappears if one considers quantum corrections. It will be an important future problem to study the quantum corrections to $E_W$ using the prescription proposed in \cite{Faulkner:2013ana}. We expect that considering this quantum corrections change the phase diagram of $E_W$ and especially one find a smooth behavior near the critical point. 

Finally it would be interesting to study $E_W$ in more general holographic setups, e.g., higher curvature gravities. It is known that for such theories the RT prescription for computing HEE fails and one should use other proposals\cite{Hung:2011xb,Dong:2013qoa,Camps:2013zua,Mozaffar:2016hmg}. At present, our preliminary analysis suggests that in this case one should replace eq.\eqref{heop} with another functional which contains higher curvature corrections. We leave the details of this interesting problem for future study \cite{toappear}.

%\section{Finite Temprature}
%
%metric
%
%$d>1$:
%
%case 1: $T=0$
%
%\bea
%I_{T=0}=\frac{\mathcal{C}_0}{4G_N} H^{d-1}\left(\frac{2}{\ell^{d-1}}-\frac{1}{h^{d-1}}-\frac{1}{(2\ell+h)^{d-1}}\right)\nonumber\\
%E_{T=0}=\frac{\mathcal{C}_0}{4cG_N} H^{d-1}\left(\frac{1}{(2\ell+h)^{d-1}}-\frac{1}{h^{d-1}}\right)
%\eea
%
%
%
%case 2: $\ell T\ll 1$ and $h T\ll 1$
%
%\bea
%I=I_{T=0}-\frac{\mathcal{C}_0\mathcal{C}_1}{2G_N} H^{d-1}(\ell+h)^2T^{d+1}\nonumber\\
%E=E_{T=0}-2\frac{\mathcal{C}_0\mathcal{C}_1}{dcG_N} H^{d-1}\ell(\ell+h)T^{d+1}
%\eea
%
%
%
%case 3: $h T\ll 1\ll \ell T$
%
%
%\bea
%I=\frac{ H^{d-1}T^{d-1}}{4G_N}\left(-\frac{\mathcal{C}_0}{(hT)^{d-1}}+\left(\frac{4\pi}{d+1}\right)^{d-1}\mathcal{C}_2-\left(\frac{4\pi}{d+1}\right)^dhT-\mathcal{C}_0\mathcal{C}_1(hT)^2\right)
%\eea

\appendix
%\section{????}

%The entropy of purification
%\bea
%E_P=\frac{L^dH^{d-1}}{4G_N}\int_{r_t^a}^{r_t^b}\frac{dr}{r^{d}}=\frac{L^dH^{d-1}}{4G_N}\frac{c^{d-1}}{d-1}\left(\frac{1}{a^{d-1}}-\frac{1}{b^{d-1}}\right).
%\eea
%
%
%
%On the other hand
%\bea
%\frac{1}{2}I=\frac{L^dH^{d-1}c^{d}}{4G_N(d-1)}\left(\frac{1}{a^{d-1}}+\frac{1}{b^{d-1}}-\frac{2^d}{(b-a)^{d-1}}\right).
%\eea
%So
%\bea\label{ineqEI}
%E_P\geq \frac{1}{2}I\rightarrow \frac{1}{a^{d-1}}-\frac{1}{b^{d-1}}\geq c\left(\frac{1}{a^{d-1}}+\frac{1}{b^{d-1}}-\frac{2^d}{(b-a)^{d-1}}\right)
%\eea
%We have
%\bea
%b+a\geq 0\rightarrow \frac{2}{b-a}\geq \frac{1}{b}\rightarrow\frac{2^d}{(b-a)^{d-1}}\geq \frac{2}{b^{d-1}}\rightarrow\nonumber\\
%\frac{1}{a^{d-1}}-\frac{1}{b^{d-1}}\geq \frac{1}{a^{d-1}}+\frac{1}{b^{d-1}}-\frac{2^d}{(b-a)^{d-1}}
%\eea
%$c\leq 1$ so Eq.\eqref{ineqEI} holds.
%\begin{tikzpicture}
%      \draw[->] (-3,0) -- (4.2,0) node[right] {$x$};
%      \draw[->] (0,-3) -- (0,4.2) node[above] {$y$};
%      \draw[scale=0.5,domain=-1:1,smooth,variable=\x,blue] plot ({\x},{\x*\x*\x*\x});
%    \end{tikzpicture}

\subsection*{Acknowledgements}
We are very grateful to Mohsen Alishahiha and Ali Mollabashi for correspondence, careful
reading of the manuscript and their valuable comments.

\section{Thermal Corrections to HEE and HMI in Nonrelativistic Theories}\label{LHTNR}
In this appendix we briefly review the low and high temperature expansions of HEE and HMI in nonrelativistic theories with Lifshitz and hyperscaling violating exponents. Here we only focus on the main steps and for further discussions we refer to \cite{Fischler:2012uv}.

Similar to section \ref{sec:LHT} we consider a strip entangling region (see Fig. \ref{fig:regions}) for computing HEE and HMI and parametrize the corresponding hypersurface as eq.\eqref{stripregion}. Employing the RT prescription and using eq.\eqref{metricNR}, the corresponding HEE functional is given by
\bea\label{RTstripNR}
S=\frac{H^{d-1}}{4G_N \,\rft}\int\frac{dr}{r^{\dt}}\sqrt{\frac{1}{f(r)}+x'(r)^2},
\eea
where we have defined an effective dimension $\dt=d-\theta$. Extremizing this functional yields the equation of motion for $x(r)$
\bea\label{eomtempNR}
x'(r)=\pm\frac{1}{\sqrt{f(r)\left(\left(\frac{r_t}{r}\right)^{2\dt}-1\right)}},
\eea
where $r_t$ denotes the turning point of the minimal hypersurface. Now we can obtain the relation between $\ell$ and $r_t$  as
\bea\label{lengthtempNR}
\ell=2r_t\int_0^1\frac{u^\dt du}{\sqrt{1-u^{2\dt}}}\left(1-\left(\frac{r_t}{r_0}\right)^{\dt+z} u^{\dt+z}\right)^{-\frac{1}{2}}.
\eea
In addition, the HEE is given by on-shell functional of eq.\eqref{RTstripNR} for the hypersurface eq.\eqref{eomtempNR} 
\bea\label{heetempNR}
S=\frac{1}{2G_N\,\rft}\frac{H^{d-1}}{r_t^{\dt-1}}\int_{\frac{\epsilon}{r_t}}^{1}\frac{du \, u^{-\dt}}{\sqrt{1-u^{2\dt}}}\left(1-\left(\frac{r_t}{r_0}\right)^{\dt+z} u^{\dt+z}\right)^{-\frac{1}{2}}.
\eea
%\subsection{Special Cases}
One may consider two special cases $d=\theta$ and $\theta=d-1, \; z=1$. In the former, the RT surface lies on the boundary slice $r=\epsilon$ and we have an extensive violation of area law \cite{Dong:2012se}.  The latter, up to an overall factor $H^{\theta}$, is exactly same as $d=1$ relativistic theory so we do not mention it again (see section \ref{d=1}). Therefore, in the rest, we neglect these two special cases.
%\subsubsection{HEE and HMI in $d \neq \theta$}

By employing method of section \ref{sec:finiteT} we can extract the behavior of HEE and HMI at low and high temperature. Using binomial series and performing the integral  we obtain the relation between length of entangling region $\ell$ and turning point $r_t$ as
\bea\label{ellrtNR}
\ell=2r_t\sum_{n=0}^{\infty}\frac{1}{1+n(\dt+z)}\frac{\Gamma\left(n+\frac{1}{2}\right)}{\Gamma\left(n+1\right)}\frac{\Gamma\left(\frac{\dt+1+n(\dt+z)}{2\dt}\right)}{\Gamma\left(\frac{(\dt+z)n+1}{2\dt}\right)}\left(\frac{r_t}{r_0}\right)^{n(\dt+z)},
\eea
This series converges for $r_t<r_0$ when $\dt+z>0$. A similar calculation for HEE functional eq.\eqref{heetempNR} shows that
%\bea
%S=\frac{1}{2G_N(\dt-1)}\qty(\frac{r_f}{\epsilon})^{\theta}\left(\frac{H}{\epsilon}\right)^{d-1}+S_{\rm finite.},
%\eea
\bea
S=\frac{1}{2G_N(\dt-1) }\frac{H^{d-1}}{\rft}\frac{1}{\epsilon^{\dt-1}}+S_{\rm finite.},
\eea
where
\bea\label{SfiniteNR}
S_{\rm finite.}=\frac{1}{4G_N}\frac{H^{d-1}}{\rft \, r_t^{\dt-1}}\left(\frac{\ct}{1-\dt}+\sum_{n=1}^{\infty}\frac{1}{\dt}\frac{\Gamma\left(n+\frac{1}{2}\right)}{\Gamma\left(n+1\right)}\frac{\Gamma\left(\frac{n(\dt+z)-\dt+1}{2\dt}\right)}{\Gamma\left(\frac{n(\dt+z)+1}{2\dt}\right)}\left(\frac{r_t}{r_0}\right)^{n(\dt+z)}\right),
\eea
and we have define $\ct=\frac{2\sqrt{\pi}\,\Gamma \left(\frac{\dt+1}{2\dt}\right)}{\Gamma \left(\frac{1}{2\dt}\right)}$. 
Using these results we can find the low and high temperature behavior of HEE and HMI. 

\subsubsection*{(i) HEE at Low Temperature Limit $\ell T^{\frac{1}{z}}\ll 1$}
By using eq.\eqref{tempNR} the $\ell T^{\frac{1}{z}}\ll 1$ limit can be interpreted as $r_t\ll r_0$. In this limit we can solve eq.\eqref{ellrt} for $r_t$  
\bea\label{rtsmalltempNR}
r_t=\frac{\ell}{\ct}\left(1-\frac{\sqrt{\pi}}{(\dt+z+1)\ct^{\dt+z+1}}\frac{\Gamma\left(\frac{2\dt+z+1}{2 \dt}\right)}{\Gamma\left(\frac{\dt+z+1}{2\dt}\right)}\left(\frac{\ell}{r_0}\right)^{\dt+z}+\mathcal{O}\left(\left(\frac{\ell}{r_0}\right)^{2(\dt+z)}\right)\right).
\eea
Plugging $r_t$ into eq.\eqref{SfiniteNR} and using eq.\eqref{tempNR}, the low temperature corrections to finite part of HEE obtains
\bea\label{sfinitelowtempNR}
S_{\rm finite.}=\frac{\Ct_0}{4G_N\,\rft} \frac{H^{d-1}}{\ell^{d-1}}\left(1+\Ct_1 (\ell T^{\frac{1}{z}})^{\dt+z}+\mathcal{O}((\ell T)^{2(\dt+z)})\right),
\eea
where 
\bea \label{C0C1}
\Ct_0= \frac{\ct^\dt}{1-\dt},\;\;\;\;\;\;
\Ct_1=\left(\frac{4\pi}{\vert\dt+z\vert}\right)^{\frac{\dt+z}{z}}\frac{\sqrt{\pi}}{2\ct^{\dt+z+1}}\frac{1-\dt}{\dt+z+1}\frac{\Gamma \left(\frac{z+1}{2\dt}\right)}{\Gamma \left(\frac{\dt+z+1}{2\dt}\right)}.
\eea
For  $z>-1-\dt$ the thermal correction increases the HEE. However, for $z<-1-\dt$ the HEE decreases by thermal correction, but as we mentioned, the negative value of $z$ (for $\theta>d$) has been excluded by thermodynamic stability of black brane solution.
\subsubsection*{(ii) HEE at High Temperature Limit $\ell T\gg 1$}
In the high temperature the near horizon part of RT surface has the main  contribution to HEE. Therefore, to obtain HEE we can consider eqs.\eqref{ellrtNR} and \eqref{SfiniteNR} in the limit that $r_t \rightarrow r_0$. By manipulating eq.\eqref{SfiniteNR} we get 
\begin{align}
S_{\rm finite}&=\frac{H^{d-1}}{4G_N \, \rft\, r_t^{\dt-1}}\Bigg{(}\frac{\ell}{r_t}-\frac{\ct\dt}{\dt-1}+ \sum_{n=1}^{\infty}\frac{1}{2\dt}\frac{\Gamma \left(n+\frac{1}{2}\right)}{\Gamma (n+1)}\frac{\Gamma\left(\frac{n(\dt+z)-\dt+1}{2\dt}\right)}{\Gamma\left(\frac{n(\dt+z)+2\dt+1}{2\dt}\right)}\left(\frac{r_t}{r_0}\right)^{n(\dt+z)}\Bigg{)}.\nonumber
\end{align}
In the large $n$ limit the above series behaves as $\frac{1}{n^2}\left(\frac{r_t}{r_0}\right)^{n\dt+z}$, so we can take  $r_t\rightarrow r_0$ limit. Now using eq.\eqref{tempNR}, we obtain the finite part of HEE at high temperature
\bea\label{sfinitehightempNR}
S_{\rm finite}=\frac{\ell H^{d-1}}{4G_N \, \rft}\left(\frac{4\pi T}{\vert \dt+z \vert}\right)^{\frac{\dt}{z}}\left(1+\qty(\frac{\vert \dt+z \vert}{4\pi\, \ell^z \, T})^{\frac{1}{z}} \Ct_2\right),
\eea
where
\bea
\Ct_2=-\frac{\ct\dt}{\dt-1}+ \sum_{n=1}^{\infty}\frac{1}{2\dt}\frac{\Gamma \left(n+\frac{1}{2}\right)}{\Gamma (n+1)}\frac{\Gamma\left(\frac{n(\dt+z)-\dt+1}{2\dt}\right)}{\Gamma\left(\frac{n(\dt+z)+2\dt+1}{2\dt}\right)}.
\eea

\subsubsection*{(iii) HMI at Low and High Temperature Limit}
Now we are ready to obtain the HMI for the configuration depicted in Fig. \ref{fig:regions} using eqs.\eqref{sfinitelowtempNR} and \eqref{sfinitehightempNR}. We only consider cases where the HMI is non-zero corresponding to a connected configuration. In this case using eq.\eqref{HMId} for eq.\eqref{sfinitelowtempNR} and assuming $h T^{\frac{1}{z}}\ll \ell T^{\frac{1}{z}}\ll 1$ we get the HMI
\bea\label{hmiexpand1NR}
I=I_{T=0}-\Ct_0\Ct_1\frac{H^{d-1}}{2G_N\,\rft}\;T^\frac{{\dt+z}}{z}  \qty((h+2\ell)^{z+1}-2\ell^{z+1}+h^{z+1}),
\eea
where $I_{T=0}$ is the HMI at zero temperature
\bea
I_{T=0}=\Ct_0\frac{ H^{d-1}}{4G_N\, \rft}\left(\frac{2}{\ell^{\dt-1}}-\frac{1}{h^{\dt-1}}-\frac{1}{(2\ell+h)^{\dt-1}}\right).
\eea
Note that the subadditivity implies $\Ct_0<0$ which happens for $\dt>1$. In addition, it shows that the thermal fluctuation decreases HMI. 

Finally using eqs.\eqref{sfinitehightempNR} and \eqref{sfinitelowtempNR} for  $h T^{\frac{1}{z}}\ll 1\ll \ell T^{\frac{1}{z}}$ we can find the 
high temperature behavior of HMI  as

\bea\label{hmiexpand3}
I=\frac{ H^{d-1}T^\frac{{\dt-1}}{z}}{4G_N\,\rft}\left(-\frac{\Ct_0}{(hT^{\frac{1}{z}})^{\dt-1}}+\left(\frac{4\pi}{\vert\dt+z\vert}\right)^{\frac{\dt-1}{z}}\Ct_2-\left(\frac{4\pi}{\vert\dt+z\vert}\right)^{\frac{\dt}{z}} h T^{\frac{1}{z}}-\Ct_0\Ct_1(hT^{\frac{1}{z}})^{z+1}\right).
\eea

\end{document}